\documentclass[prl,amsmath,amssymb,twocolumn,superscriptaddress,aps,floatfix,longbibliography]{revtex4-2}
\usepackage{subfigure}
\usepackage{graphicx}
\usepackage{dcolumn}
\usepackage{bm}
\usepackage{xcolor}
\usepackage{float}
\usepackage{standalone}
\usepackage{comment}
\usepackage[hidelinks]{hyperref}

 \graphicspath{ {./images/} }
 \usepackage{physics}

\begin{document}

\preprint{APS/123-QED}

\title{Chiral phonons induced from spin dynamics via magnetoelastic anisotropy}

\author{Bowen Ma}
\email{bowenphy@hku.hk}
\affiliation{Department of Physics and HK Institute of Quantum Science \& Technology, 
The University of Hong Kong, Pokfulam Road, Hong Kong, China}

\author{Z. D. Wang}
\email{zwang@hku.hk}
\affiliation{Department of Physics and HK Institute of Quantum Science \& Technology, 
The University of Hong Kong, Pokfulam Road, Hong Kong, China}
\affiliation{Quantum Science Center of Guangdong-Hong Kong-Macau Great Bay Area, 3 Binlang Road, Shenzhen, China}

\author{Gang V. Chen}
\email{chenxray@pku.edu.cn}
\affiliation{International Center for Quantum Materials, 
School of Physics, Peking University, Beijing 100871, China}

\date{\today}

\begin{abstract}
We propose a mechanism to obtain chiral phonon-like excitations from the bond-dependent magnetoelastic couplings in the absence of out-of-plane magnetization and magnetic fields. By mapping the hybrid excitation to its phononic analog, we reveal the impact of the lattice symmetry on the origin of the chirality. In the example of a triangular lattice ferromagnet, we recognize that the system is equivalent to the class D of topological phonons, and show the tunable chirality and topology by an in-plane magnetic field. As a possible experimental probe, we evaluate the phonon magnetization and planar thermal Hall conductivity. Our study gives a new perspective on tunable topological and chiral excitations beyond the Raman spin-lattice coupling, suggesting possible applications of spintronics and phononics in various anisotropic magnets and/or Kitaev materials.
\end{abstract}

\maketitle

\emph{\textcolor{blue}{Introduction.}}---In recent years, chiral phonons,
i.e. circularly polarized lattice vibrations with non-zero angular momentum, have attracted great attention 
to both theoretical studies~\cite{zhang2015chiral,chen2018chiral,gao2018nondegenerate,chen2019chiral,saparov2022lattice,zhang2022chiral} 
and experimental measurements~\cite{zhu2018observation,nakane2018angular,chen2019entanglement,baydin2022magnetic,ishito2023truly,ueda2023chiral}, including phonon contribution to Einstein–de Haas 
effect~\cite{zhang2014angular,mentink2019quantum,dornes2019ultrafast}, 
phononic magnetism~\cite{ren2021phonon,xiong2022effective,fransson2023chiral,hernandez2023observation,chaudhary2023giant,wu2023fluctuation}, optically driven 
chiral phono-magnetic effects~\cite{juraschek2019orbital,juraschek2020phono,juraschek2022giant}. 
These raise great interest in utilizing phonons for information storage 
and transport~\cite{wang2008phononics,wang2008thermal}.

In most cases, the circularly polarized right- and left-handed phonon modes are degenerated due to the inversional and time-reversal symmetry of the lattice vibrations. To lift the degeneracy, the original proposal of chiral phonons~\cite{zhang2015chiral} is to introduce Raman-type spin-lattice coupling (RSLC)~\cite{sheng2006theory,ray1967dynamical,ioselevich1995strongly,penc2004half} via an out-of-plane magnetic field or magnetization as an effective field. In principle, in-plane orders also break the time-reversal symmetry as well as the chiral symmetry, but this case is less studied since the orders are orthogonal to the phonon angular momentum (PAM) and thus do not contribute to the RSLC. However, recent planar thermal Hall signals on CoTiO$_3$~\cite{JMH} indicate the possible existence of non-degenerate chiral phonons in magnetic materials with in-plane collinear magnetic orders.

Besides the RSLC mechanism, it was pointed out that the phonon band with opposite chirality is split by the coupling between translational displacements and micro-rotations in non-centrosymmetric micropolar crystals~\cite{kishine2020chirality}. From this picture, the precession of the magnetic moments in magnetic systems can be regarded as extra micro-rotational degrees of freedom. A recent first-principles study also suggested that the slow dynamics of the spins in magnetic materials should be treated on equal footing with phonons~\cite{bonini2023frequency}. Therefore, it is very natural to ask if the dynamic effects of in-plane moments can induce chiral phonons beyond the static RSLC.

To answer this question, in this Letter, we study the lattice symmetry-allowed magnetoelastic coupling (MEC) that can naturally hybridize lattice and spin dynamics. Previous studies on magnetoelastic coupled systems~\cite{takahashi2016berry,zhang2019thermal,thingstad2019chiral,go2019topological,park2019topological,sheikhi2021hybrid,efimkin2021topological,ma2022antiferromagnetic,bao2020evidence,zhang2021anomalous,luo2023evidence} have not investigated the anisotropic effects of the lattice on the excitation's chirality. With this magnetoelastic anisotropy, we find the hybridized excitations have non-trivial charility and topology in the example of a simple triangular ferromagnet with D$_3$ point group symmetry. Nevertheless, in the case with D$_{3h}$ symmetry, though the hybridization still exists, the excitations are topologically trivial with no chirality. To understand this lattice symmetry significance, we map the spin-wave dynamics into artificial vibration modes and the full Hamiltonian can then be treated as a pure phononic Hamiltonian. With a proper lattice symmetry, a dynamical magnetic field can emerge from the MEC in the phononic Hamiltonian, leading to non-degenerated chirality and the chiral symmetry broken class D of the phonon topological classification~\cite{susstrunk2016classification}. We also calculate the planar thermal Hall conductivity based on our mechanism, which possibly explains the results measured on CoTiO$_3$~\cite{JMH}. Our work not only provides an alternative mechanism for chiral phonons in magnetic materials, but also reveals the significance of lattice symmetry on the topology of magnetoelastic coupled systems.

\begin{figure}
\includegraphics[width=8cm]{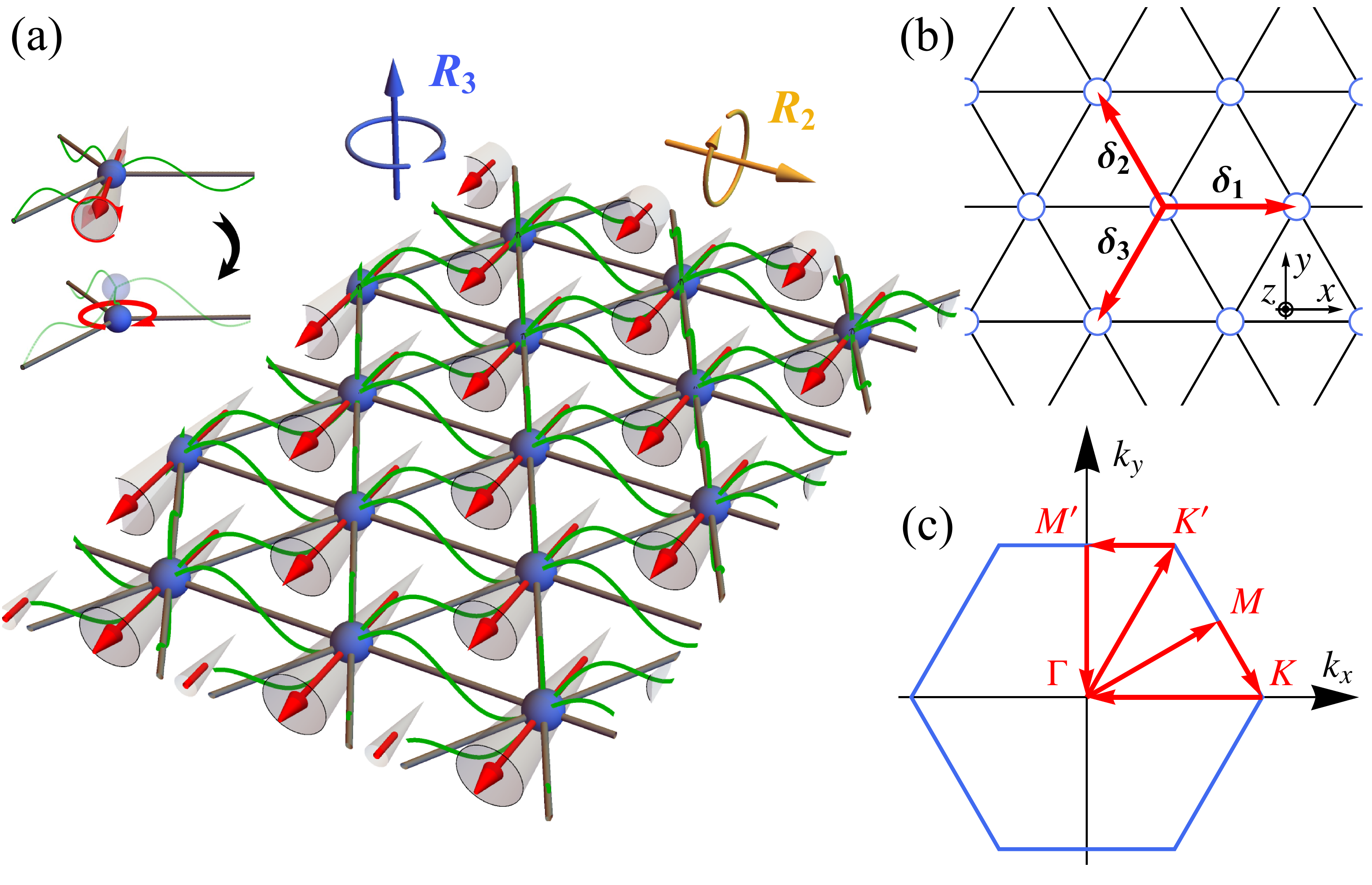}
\caption{(a) A magnetoelastic coupled system with D$_3$ lattice symmetry. The inset shows the in-plane vibration modes with spin-waves can be mapped into a pure vibration with both in-plane circular and out-of-plane modes. (b) The triangular lattice with three primary bonds $\boldsymbol{\delta}_i=(\cos 2\pi\frac{i-1}{3},\sin 2\pi\frac{i-1}{3})$. (c) The hexagonal Brillouin zone with high-symmetry path.}
\label{Schematic}
\end{figure}

\emph{\textcolor{blue}{Model.}}---We begin from a two-dimensional lattice model with in-plane magnetic moments. For the simplest case with the in-plane ferromagnetic order, we invoke a minimal \textit{bond-independent} XXZ-type spin Hamiltonian as,
\begin{align}
    H_s&=J\sum_{\langle ij\rangle}(S^x_i S^x_j+S^y_i S^y_j+\Delta S^z_i S^z_j)\nonumber\\
    &-\eta_y\sum_{i}(S_i^y)^2-\mathcal{B}\sum_{i}S_i^y,
\end{align}
where $J<0$ is the ferromagnetic nearest-neighbor (N.N.) exchange coupling, and $0<\Delta<1$ is the easy-plane XXZ anisotropy. With the lower symmetry in the ferromagnetic state, 
the material would in principle have additional interactions 
via the spin-lattice coupling~\cite{penc2004half} or quantum order by disorder mechanism~\cite{maksimov2019anisotropic} to further stabilize the orders. Hence, without the loss of generality, we add a weak single-ion anisotropy $\eta_y$ 
and a Zeemann splitting $\mathcal{B}=g\mu_B B$ along the $y$-direction 
to force the moments along the $y$-axis. To obtain the spin dynamics, We write the spin operators into the standard Holstein-Primakoff representation~\cite{holstein1940field} as $S_i^\pm=S_i^z\pm iS_i^x\approx \sqrt{2S}a^{}_i (a^\dagger_i)$, and $S_i^y = S-a^\dagger_ia^{}_i$, where $a^{}_i (a^\dagger_i)$ is the magnon annihilation (creation) operator at site $\mathbf{r}_i$. Then, Eq.(2) can be Fourier transformed into a Bogoliubov-de Gennes form as
\begin{align}
    H_s&=\frac{1}{2}\sum_\mathbf{k}(a^\dagger_\mathbf{k},a^{}_{-\mathbf{k}})H_m(\mathbf{k})\begin{pmatrix}
    a^{}_\mathbf{k} \\ a^\dagger_{-\mathbf{k}}
\end{pmatrix}\nonumber\\
&=\frac{1}{2}\sum_\mathbf{k}(a^\dagger_\mathbf{k},a^{}_{-\mathbf{k}})\begin{pmatrix}
    A_m(\mathbf{k}) & B_m(\mathbf{k})\\
    B^*_m(-\mathbf{k}) & A^*_m(-\mathbf{k})
\end{pmatrix}\begin{pmatrix}
    a^{}_\mathbf{k} \\ a^\dagger_{-\mathbf{k}}
\end{pmatrix}
\end{align}
with magnon dispersion $\epsilon_m(\mathbf{k})=\sqrt{A_{m\mathbf{k}}^2-B_{m\mathbf{k}}^2}$, where $A_{m}(\mathbf{k})=B_\text{eff}-zJS+\frac{1}{2}JS(1+\Delta)\sum_{i\in \text{N.N.}} e^{i\mathbf{k}\cdot\mathbf{r}_i}$ and $B_m(\mathbf{k})=-\frac{1}{2}JS(1-\Delta)\sum_{i\in \text{N.N.}} e^{i\mathbf{k}\cdot\mathbf{r}_i}$ with $B_\text{eff}\equiv\mathcal{B}-(1-2S)\eta_y$ and $z$ is the coordinate number.

The lattice dynamics are described by the common elastic Hamiltonian 
\begin{align}
    H_p=\sum_{i}\frac{\mathbf{p}^2_i}{2M}
    +\frac{M\omega_0^2}{2}\sum_{\langle ij \rangle}(\hat{\mathbf{R}}_{ij}^0\cdot\mathbf{u}_{ij})^2,
    \label{PhononH}
\end{align}
where ${\mathbf{u}_{ij}=\mathbf{u}_j-\mathbf{u}_i}$ is the in-plane displacement 
of the lattice, $\mathbf{p}_i$ is the corresponding momentum, 
$\hat{\mathbf{R}}_{ij}^0$ is the unit vector along bond $ij$ in equilibrium, 
$M$ is the atom mass and $\omega_0$ is the intrinsic vibration frequency. For single-atom lattices, Eq.~(1) can be written as $H_p=\sum_\mathbf{k}\frac{1}{2M}|\mathbf{p}_{\mathbf{k}}|^2+\frac{1}{2}\mathbf{u}_{\mathbf{k}}^\dagger D_p(\mathbf{k})\mathbf{u}_{\mathbf{k}}$ in the Fourier space, where $D_p(\mathbf{k})$ is the $2\times 2$ dynamical matrix~\cite{sm}.

The spin and lattice degrees of freedom are naturally coupled by magnetoelastic interactions, which can be generally written as
\begin{align}
    H_{me}=\sum_{\langle ij\rangle}\sum_{\alpha\beta}K_{ij}^{\alpha\beta} 
    S_{i}^\alpha S_{j}^\beta(\hat{\mathbf{R}}_{ij}^0\cdot\mathbf{u}_{ij}),
    \label{ME}
\end{align}
where $K_{ij}^{\alpha\beta}$'s are the coupling constants and they should respect the lattice symmetry~\cite{kittel1949physical,kittel1958interaction,streib2019magnon,ruckriegel2020angular,sm}. The diagonal terms such as $K_{ij}^{yy}$ can contribute to the anisotropy along $y$-direction as we mentioned before, while the off-diagonal terms can be re-expressed as
\begin{align}
    S_{i}^y (K_{ij}^{y+} 
     S_{j}^-+K_{ij}^{y-} 
     S_{j}^+)(\hat{R}_{ij}^{0+} u^-_{ij}+\hat{R}_{ij}^{0-}u^+_{ij}),
\end{align}
where $K_{ij}^{y\pm}\equiv K_{ij}^{yz}\pm i K_{ij}^{yx}$ and $\hat{R}_{ij}^{0\pm}\equiv \hat{R}_{ij}^{0x}\pm i\hat{R}_{ij}^{0y}$. Here, it is the spin dynamics $S^\pm$ that couples with the circular modes of phonon $u^\pm_{ij}=u^x_{ij}\pm iu^y_{ij}$, so that the procession of the spin-wave is possible to transfer the angular momentum to the wiggling of the lattice.

\emph{\textcolor{blue}{Chiral excitations.}}---To demonstrate that the spin dynamics can induce phonon chirality, we study a magnetoelastic coupled system on a triangular lattice with D$_3$ symmetry and evaluate the full Hamiltonian 
\begin{align}
    H=H_s+H_p+H_{me}
\end{align}
with magnons and phonons on equal footing. Because of the bond inversion symmetry, the antisymmetric interactions, such as DMI, are not allowed, and thus $K_{ij}^{\alpha\beta}=K_{ij}^{\beta\alpha}$. To respect the D$_3$ symmetry~\cite{sm,gallego2019automatic}, we have the \textit{bond-dependent} coupling strength
\begin{equation}
    \scalebox{0.9}{\text{$K_{ij}=\begin{pmatrix}
        K_0+K_1\cos 2\phi_{ij} & K_1\sin 2\phi_{ij} & K_2\sin 2\phi_{ij}\\K_1\sin 2\phi_{ij} & K_0-K_1\cos 2\phi_{ij} & K_2\cos 2\phi_{ij}\\K_2\sin 2\phi_{ij} & K_2\cos 2\phi_{ij} & K_3
    \end{pmatrix}$}},\label{MECoupling}
\end{equation}
where $K_\mu$ $(\mu=0,1,2,3)$ are four independent MEC constants, and $\phi_{ij}=0$, $2\pi/3$ or $4\pi/3$ is the $ij$-bond angle from $x$-axis to repsect the three-fold rotational symmetry $R_3$.

Different from previous studies on square lattices~\cite{kittel1949physical,kittel1958interaction}, where $K_{ij}^{\alpha\beta}$ only contain two independent constants and only the out-of-plane vibrational mode is coupled with spins, in our case the MEC couples the moments with in-plane phonon modes. With hybrid basis $\mathbf{X}_\mathbf{k}=(a^{}_\mathbf{k},a^\dagger_{-\mathbf{k}},u^x_\mathbf{k},u^y_\mathbf{k},p^x_{-\mathbf{k}},p^y_{-\mathbf{k}})^T$, the full Hamiltonian is written as $H=\frac{1}{2}\sum_{\mathbf{k}}\mathbf{X}_\mathbf{k}^\dagger H_\mathbf{k}\mathbf{X}_\mathbf{k}$ with
\begin{align}
    H_\mathbf{k}=\begin{pmatrix}
        H_m(\mathbf{k}) & H_{c}(\mathbf{k}) & 0\\
        H^\dagger_{c}(\mathbf{k}) & D_p(\mathbf{k}) & 0\\
        0 & 0 & \frac{I_2}{M}
    \end{pmatrix},
\end{align}
where the spin and lattice dynamics are coupled by
\begin{align}
    &H_{c}(\mathbf{k})=K_1\sqrt{\frac{S^3}{2}}\sum_i\begin{pmatrix}
        \sin 2\phi_i & 1-\cos 2\phi_i\\
        -\sin 2\phi_i & \cos 2\phi_i-1
    \end{pmatrix}\sin\mathbf{k}_i\nonumber\\
    &+iK_2\sqrt{\frac{S^3}{2}}\sum_i\begin{pmatrix}
        1+\cos 2\phi_i & \sin 2\phi_i\\
        1+\cos 2\phi_i & \sin 2\phi_i
    \end{pmatrix}\sin\mathbf{k}_i
\end{align}
that only involves $K_1$ and $K_2$. Here, $\mathbf{k}_i\equiv\mathbf{k}\cdot\boldsymbol{\delta}_i$ with three primary bonds $\boldsymbol{\delta}_i=(\cos 2\pi\frac{i-1}{3},\sin 2\pi\frac{i-1}{3})$ and $\phi_i=2\pi\frac{i-1}{3}$ $i=1,2,3$. Given the commutator
\begin{align}
    [\mathbf{X}^{}_\mathbf{k},\mathbf{X}^\dagger_\mathbf{k}]=\begin{pmatrix}
    \sigma_z & 0 & 0\\
    0 & 0 & iI_3\\
    0 & -iI_3 & 0
\end{pmatrix}\equiv \mathcal{J},
\end{align}
one can derive the eigen-equation for $n$-th wavefunction $|\psi_{n\mathbf{k}}\rangle$ as $E_{n\mathbf{k}}|\psi_{n\mathbf{k}}\rangle=
\mathcal{J} H_{\mathbf{k}}|\psi_{n\mathbf{k}}\rangle$ by Heisenberg equation of motion.

\begin{figure}
\includegraphics[width=0.46\textwidth]{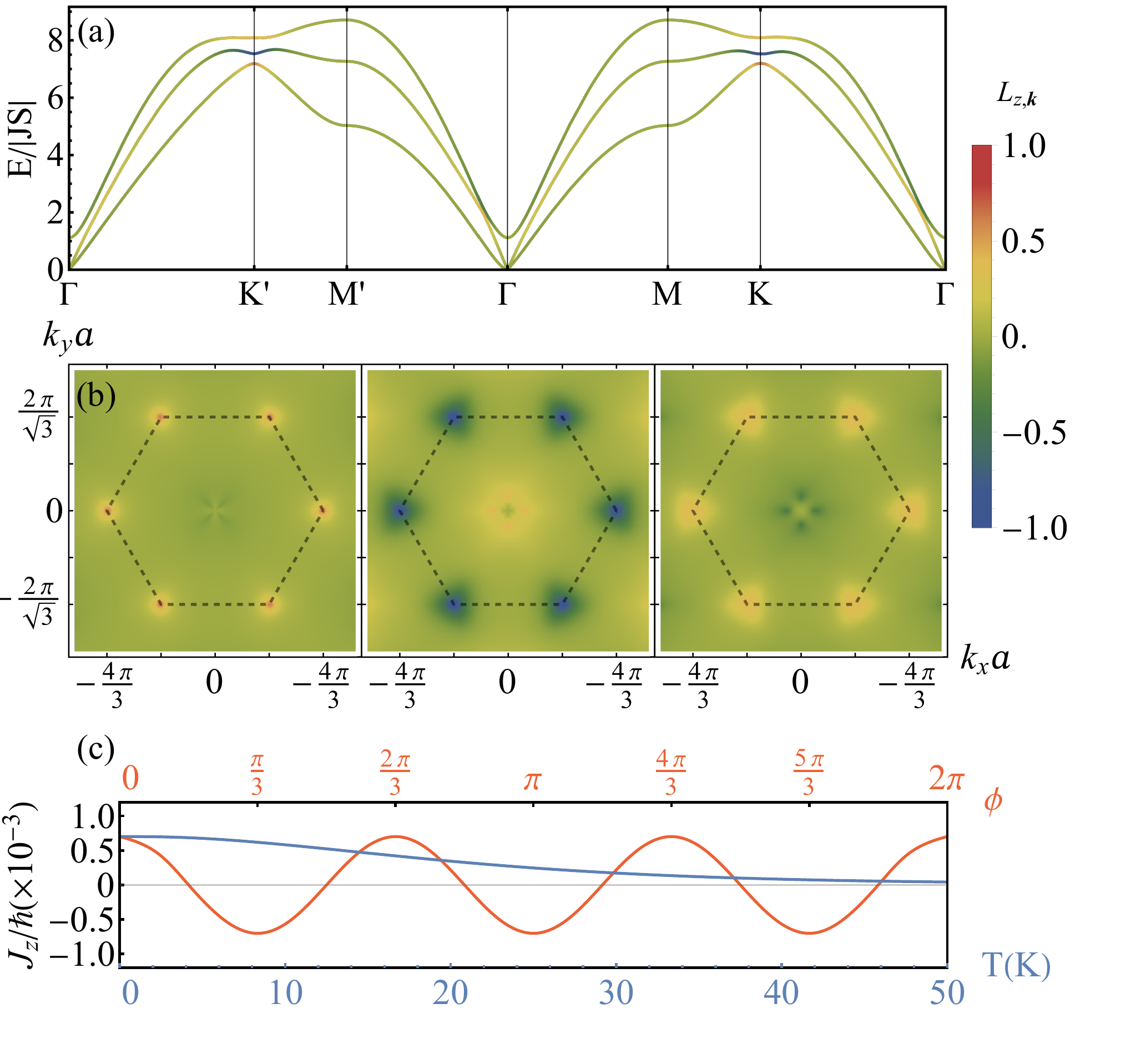}
    \caption{Chiral excitations of magnon-phonon hybridization with ${S=3/2}$, ${J=-1.5}$ meV, ${\Delta=0.1}$, ${\hbar\omega_0=8}$ meV, 
$M = 56$ u, $K_1=K_2=6$ meV/\AA, and ${B_\text{eff}=1}$ meV. (a) The band dispersion with anti-crossing gaps. (b) The finite angular momentum distribution in the momentum space $L_{z,\mathbf{k}}$ from left to right for the bottom, middle, and top bands respectively. (c) The dependence of PAM per atom on temperature (blue curve) and on magnetic order direction $\phi$ from $y$-axis (orange curve).} 
    \label{fig:chirality}
\end{figure}

In a pure phononic system, the chiral phonon angular momentum is defined as ${L_z=\sum_i\mathbf{u}_i\times\mathbf{p}_i}$~\cite{zhang2014angular,chen2018chiral,chen2019chiral}. In our magnon-phonon hybrid system, since the magnetic moment is along the $y$-direction, 
it will not obfuscate the physical meaning of this definition along the $z$-direction. In Fig.~\ref{fig:chirality}, with a typical set of parameters (as specified in the figure caption) 
for $3d$ transition-metal oxides, we depict the dispersion $E_{n\mathbf{k}}$ colored by the momentum-resolved PAM $L_{z,n\mathbf{k}}=\langle\psi_{n\mathbf{k}}|\mathbf{u}_\mathbf{k}\times\mathbf{p}_{-\mathbf{k}}|\psi_{n\mathbf{k}}\rangle$. Here, we take ${K_{1}=K_{2}=6}$ meV/ \AA, 
${S=3/2}$ and the lattice constant ${a= 5}$ \AA,    
which is in the same order as the estimate $J\ll Ka\ll 100J$ in Ref.~\cite{streib2019magnon}.
We find that there are anti-crossings between the magnon and phonon bands and the degeneracy at $\mathbf{K}$ and $\mathbf{K}'$ of the 
original two phonon bands is lifted. Non-zero angular momenta are concentrated at these opened gaps at $\mathbf{K}$, $\mathbf{K}'$, 
and those magnon-phonon anticrossings.

The opposite chirality can in principle be detected by circular dichroism~\cite{zhu2018observation}, and additionally a net magnetization from chiral excitations~\cite{zhang2014angular,chen2018chiral,juraschek2019orbital,chaudhary2023giant} 
can be experimentally measured as
\begin{align}
    M_z^\text{ph}(T)\propto\sum_{n\mathbf{k}}L_{z,n\mathbf{k}}\left[g(E_{n\mathbf{k}},T)+\frac{1}{2}\right]\equiv J_z(T),
\end{align} 
where ${g(x,T)=[\text{exp}(x/k_B T)-1]^{-1}}$ is the Bose-Einstein distribution. 
In Fig.~\ref{fig:chirality}(c), we plot the PAM $J_z$ per atom
with respect to the temperature change as the blue curve. 
At high temperatures, it goes to zero as expected~\cite{zhang2014angular}. 
We point up that this non-zero PAM can be observed in equilibrium 
even without external magnetic field $\mathcal{B}$. 

Remarkably, the magnetic field can indirectly tune this phonon magnetization 
by rotating the magnetic moments. If we set ${\eta_y=0}$ to make 
the moment orientation tunable, and assume the ferromagnetic order 
is following the direction of the magnetic field $\mathcal{B}$, 
we obtain the changing of the phonon chirality at zero temperature limit 
shown as the orange curve in Fig.~\ref{fig:chirality}(c). 
The $\frac{2\pi}{3}$-periodicity can be explained by the $R_3$ 
rotational symmetry of the Hamiltonian when ${\eta_y=0}$, 
and the chirality flipping with the reversed order can 
be readily understood by time-reversal operation or the $R_{2x}$ rotation.


\emph{\textcolor{blue}{Lattice symmetry and origin of the chirality.}}---As a sharp contrast, we investigate the same Hamiltonian but with D$_{3h}$ symmetry so that $K_2=0$ is enforced. In this case, magnons and phonons remain coupled by $K_1$ terms so that the degeneracy at valleys is still lifted. However, we find that no chiral phonon exists~\cite{sm}.

To better understand the origin of the phonon chirality from the MEC, we investigate an equivalent phononic system by mapping magnon operators into artificial lattice degrees of freedom as
\begin{align}
    u^z_\mathbf{k}=\frac{1}{2r_\mathbf{k}}(a^{}_\mathbf{k}-a^\dagger_{-\mathbf{k}}),\quad p^z_{\mathbf{k}}=i r_\mathbf{k}(a^\dagger_\mathbf{k}+a^{}_{-\mathbf{k}}),\label{Map}
\end{align} so that the canonical commutator $[u^z_\mathbf{k},p^z_{\mathbf{k}'}]=i\delta_{\mathbf{k}\mathbf{k}'}$ is satisfied, and we choose $r^2_\mathbf{k}=\frac{M}{2}\left[A_m(\mathbf{k})+B_m(\mathbf{k})\right]$ to renormalize the magnon Hamiltonian as $H_m=\frac{1}{2M}\sum_\mathbf{k} 
\left[|p^z_{\mathbf{k}}|^2+M^2\varepsilon^2_m(\mathbf{k})|u^z_\mathbf{k}|^2\right]$. Then under the ``phase space'' coordinate $(\mathbf{Q}_{\mathbf{k}}; 
\mathbf{P}^\dagger_{\mathbf{k}})=(\mathbf{u}_\mathbf{k},u^z_\mathbf{k};\mathbf{p}^\dagger_{\mathbf{k}},{p^z_{\mathbf{k}}}^\dagger)$, the full Hamiltonian can be expressed in a phonon-like form as
\begin{align}
    H=\frac{1}{2}\sum_{\mathbf{k}}(\mathbf{Q}^\dagger_{\mathbf{k}},\mathbf{P}_{\mathbf{k}}) \begin{pmatrix}
        \mathbf{D}_\mathbf{k} & \mathbf{A}_\mathbf{k}\\
        \mathbf{A}_\mathbf{k}^\dagger & \frac{I_3}{M}
    \end{pmatrix}\begin{pmatrix}
        \mathbf{Q}_{\mathbf{k}}\\\mathbf{P}^\dagger_{\mathbf{k}}
    \end{pmatrix},
    \label{PseudoPhononH}
\end{align}
with
\begin{align}
    \mathbf{D}_\mathbf{k}\!=\!\begin{pmatrix}
        \mathbf{D}_p(\mathbf{k}) & \mathbf{D}_{me}(\mathbf{k})\\
        \mathbf{D}^T_{me}(\mathbf{k}) & M\varepsilon^2_{m}(\mathbf{k})
    \end{pmatrix},\ \mathbf{A}_\mathbf{k}\!=\!\begin{pmatrix}
        0 & 0 & \frac{B_\mathbf{k}^y}{M}\\
        0 & 0 & -\frac{B_\mathbf{k}^x}{M}\\
        0 & 0 & 0
    \end{pmatrix},
    \label{Dk}
\end{align}
where $\mathbf{D}_{me}(\mathbf{k})=2K_1r_\mathbf{k}\sqrt{2S^3}\sum_i\begin{pmatrix}
    \cos \phi_i \\ \sin \phi_i
\end{pmatrix}\sin \phi_i\sin\mathbf{k}_i$, and $\begin{pmatrix}
    B_\mathbf{k}^x \\ B_\mathbf{k}^y
\end{pmatrix}=K_2 r_\mathbf{k}^{-1}M\sqrt{2S^3}\sum_i\begin{pmatrix}
    \sin\phi_i \\ -\cos \phi_i
\end{pmatrix}\cos \phi_i\sin\mathbf{k}_i$.

From Eq.~\eqref{PseudoPhononH}, the equation of motion for the ``coordinate'' 
$\mathbf{Q}_\mathbf{k}$ can be derived as~\cite{sm}
\begin{equation} 
\ddot{\mathbf{Q}}_\mathbf{k}
+2\tilde{\mathbf{A}}_{\mathbf{k}}\dot{\mathbf{Q}}_\mathbf{k} 
+\Tilde{\mathbf{D}}_{\mathbf{k}}\mathbf{Q}_\mathbf{k}=0,
\label{PhononEoM}
\end{equation}
where  
$\tilde{\mathbf{D}}_{\mathbf{k}}=\frac{1}{M}\mathbf{D}_\mathbf{k}-\mathbf{A}_\mathbf{k}^\dagger\mathbf{A}_\mathbf{k}$  
and $\tilde{\mathbf{A}}_{\mathbf{k}}=\frac{1}{2}(\mathbf{A}_\mathbf{k}-
    \mathbf{A}_\mathbf{k}^\dagger)$.
Formally, Eq.~\eqref{PhononEoM} is exactly the same as 
the dynamical equation for a purely phononic system~\cite{saito2019berry} 
with the dynamical matrix $\tilde{\mathbf{D}}_{\mathbf{k}}$ 
under a magnetic field $\mathbf{B}_\mathbf{k}=(B_\mathbf{k}^x, B_\mathbf{k}^y, 0)$ 
as 
\begin{align}
    \tilde{H}=\sum_{\mathbf{k}}\frac{1}{2M}|\mathbf{P}_{-\mathbf{k}}-\frac{1}{2}\mathbf{B}_\mathbf{k}\times\mathbf{Q}_\mathbf{k}|^2+\frac{M}{2}\mathbf{Q}_\mathbf{k}^\dagger\tilde{\mathbf{D}}_{\mathbf{k}}\mathbf{Q}_\mathbf{k},
\end{align}
where $\frac{1}{2}\mathbf{B}_\mathbf{k}\times\mathbf{Q}_\mathbf{k}$ behaves more like a fictitious vector gauge potential and contributes to the phonon chirality similar to the RSLC mechanism. Therefore, even though the spin dynamics and lattice dynamics are coupled by $\mathbf{D}_{me}(\mathbf{k})$ with $K_1\neq 0$, the dynamical magnetic field $\mathbf{B}_\mathbf{k}$ is absent to induce chiral phonons when $K_2=0$. By investigating the lattice symmetry with TENSOR~\cite{gallego2019automatic,sm}, we find that systems with D$_6$, C$_{6v}$, D$_{3h}$ or D$_{6h}$ point group symmetry in hexagonal crystal family enforce $K_2=0$ and thus they do not possess chiral phonons from our magnetoelastic mechanism. Particularly, such chiral excitations can in principle exist in CoTiO$_3$ where the $K_2$ terms are allowed by the C$_{3i}$ point group symmetry.

\emph{\textcolor{blue}{Topological effects.}}---It is known that, when phonons couple to static magnetic fields via RSLC mechanism, they acquire non-trivial Berry curvature. With $\mathbf{A}_\mathbf{k}\neq 0$ in Eq.~\eqref{PseudoPhononH}, we expect similar topological effects in our model. In fact, from symmetry aspects, by considering the classification of topological phonons~\cite{susstrunk2016classification,sm}, we find that Hamiltonian Eq.~\eqref{PseudoPhononH} belongs to the class D with broken chiral symmetry, while the system belongs to the class of BDI with trivial topology in two dimensions when $K_2=0$. The non-trivial topology is characterized by an integer (Chern) number defined as $C_n=\frac{1}{2\pi}\sum_{\mathbf{k}}\Omega^z_{n\mathbf{k}}$ with
\begin{align}
{\Omega^z_{n\mathbf{k}}=i\hat{\mathbf{z}}\cdot\boldsymbol{\nabla}_{\mathbf{k}}\times
\langle \psi_{n\mathbf{k}}|\mathcal{J}\boldsymbol{\nabla}_\mathbf{k}|\psi_{n\mathbf{k}}\rangle},
\end{align}
the Berry curvature~\cite{berry1984quantal,zhang2010topological,takahashi2016berry} 
for $n$-th band.

\begin{figure}
\includegraphics[width=0.48\textwidth]{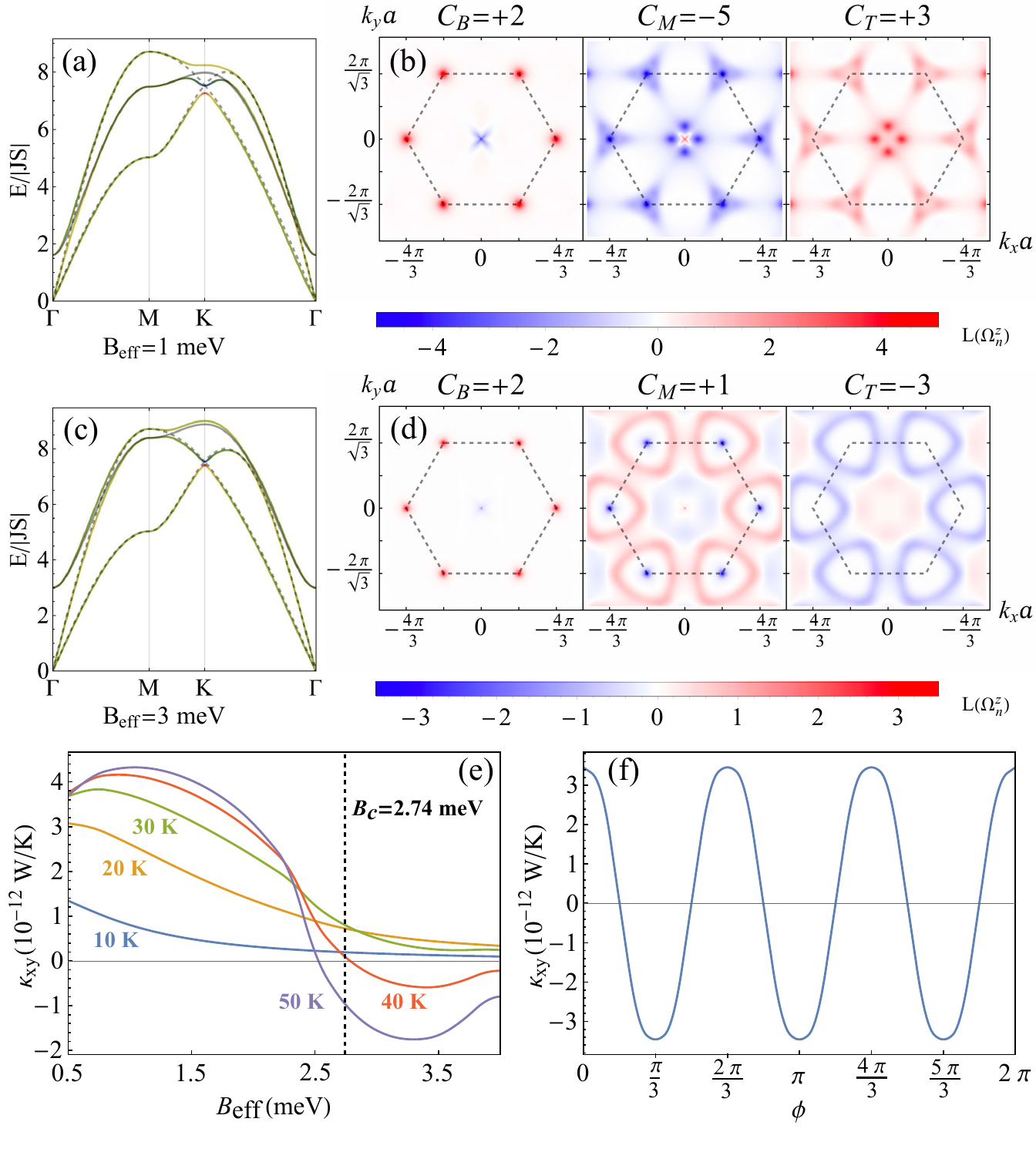}
    \caption{The distinct band topology of the chiral excitations with (a)(b) $B_\text{eff}=1$ meV and (c)(d) $3$ meV. (a)(c) The band dispersion colored by $L_{z,\mathbf{k}}$. The solid (dashed) gray line is the magnon (phonon) dispersion without MEC. (b)(d) Berry curvature distribution in log scale $L(\Omega_n^z)=\text{sgn}(\Omega_n^z)\ln(1+|\Omega_n^z|)$, and the corresponding Chern numbers $C_{B,M,T}$ for bottom, middle and top band respectively. (e) The planar thermal Hall conductivity $\kappa_{xy}$ with respect to $B_\text{eff}$ for different temperature. A sign flip around the critical effective field $B_c=2.74$ meV for $T=40$ K and $50$ K reflects the change of the middle and top band topology. (f) The dependence of $\kappa_{xy}$ at T=30 K on magnetic order direction $\phi$ from $y$-axis with $\eta_y=0$ and $B_\text{eff}=1$ meV.}
    \label{fig:Topo}
\end{figure}

We numerically evaluate the Berry curvature distribution in the momentum space and indeed find finite values around those anti-crossings 
as shown in Fig.~\ref{fig:Topo}. The degeneracy of phonons at $\mathbf{\Gamma}$ may make the Berry curvature problematic there, 
but since ${H_{mp}=0}$ at $\mathbf{\Gamma}$, $\mathbf{\Gamma}$-wavefunctions 
are topologically as trivial as the uncoupled system. Therefore, the band Chern number for each band is still well-defined. We then calculate the Chern numbers in the discretized Brillouin zone~\cite{fukui2005chern}.
With an external magnetic field along $y$-axis, one can energetically control the magnonic 
part of the band dispersion via the Zeeman effect, and thus can tune the band Chern numbers by manipulating the position of anti-crossings. 
For example, the band Chern numbers change from $(+2, -5, +3)$ to $(+2,+1,-3)$ when $B_\text{eff}>B_c=2.74$ meV, where the anti-crossing rings around $\boldsymbol{\Gamma}$ and $\mathbf{K}(\mathbf{K}')$ merge into gaps surround $\mathbf{M}(\mathbf{M}')$ along with the change of the Berry curvature distribution in the momentum space.

A direct experimental consequence of this non-trivial topology is the thermal Hall effects~\cite{strohm2005phenomenological,matsumoto2011theoretical,sugii2017thermal,grissonnanche2020chiral,zhang2024thermal}. When a longitudinal temperature gradient $\nabla_y T$ is applied to the system, 
the hybrid excitations will experience an anomalous velocity in the transverse direction, 
leading to a temperature difference in the $x$-directon. 
The corresponding thermal Hall conductivity is given 
 as~\cite{matsumoto2011rotational,matsumoto2011theoretical}
\begin{align}
\kappa_{xy}=-\frac{k_B^2T}{\hbar V}\sum_{n,\mathbf{k}}\left[c_2(g(E_{n\mathbf{k}},T))-\frac{\pi^2}{3}\right]\Omega_{n\mathbf{k}}^z,
\end{align}
where $c_2(x)=(1+x)\ln^2(1+1/x)-\ln^2x-2\text{Li}_2(-x)$, 
Li$_2(x)$ is the polylogarithm function, $T$ is the average temperature, 
$V$ is the system volume.


$\kappa_{xy}$ depends both on the Berry curvature and on its energy distribution. 
Since external fields can affect both of them, we evaluate the thermal Hall conductivity 
against $B_\text{eff}$ in Fig.~\ref{fig:Topo}(e). Here,
the magnetic field is applied in the plane and will not mix up with the conventional phonon thermal Hall effects~\cite{strohm2005phenomenological,li2020phonon,chen2022large,li2023phonon} induced by out-of-plane fields. The sign change around topological phase transition at high temperature reflects the change of middle and top band Chern numbers.
With a typical monolayer thickness about 5 \AA\ and the same parameter choice used in Fig.~\ref{fig:Topo}(e), 
an experimentally accessible thermal conductivity $\sim 10^{-3}$ WK$^{-1}$m$^{-1}$ can be achieved~\cite{onose2010observation,hirschberger2015large,banerjee2018observation}. 
With only acoustic phonons in simple triangular lattices, increasing the field will push the magnon 
dispersion away from phonon bands, and thus decrease the thermal Hall conductivity. In the case of multiple sublattices such as CoTiO$_3$, it is possible to control the magnon bands to be coupled 
with the low-energy acoustic or high-energy optical modes, giving rise to a more tunable thermal conductivity. Similar to the situation depicted in Fig~\ref{fig:chirality}(c), $\kappa_{xy}$ can also be tuned by rotating the magnetic moments with the magnetic field as shown in Fig.~\ref{fig:Topo}(f). In principle, with non-zero phonon chirality, the topological transport of spin, 
such as PAM Hall effects~\cite{park2020phonon}, 
can also occur, which we leave for future study.

\emph{\textcolor{blue}{Discussion.}}---Though we use a triangular lattice 
ferromagnet as an illustration, we remark on the general applicability of our results.
First, the magnetic order can be antiferromagnetic, non-collinear, or even paramagnetic~\cite{pocs2024generic} with or without topological magnons. For each spin $\mathbf{S}_i$,     
it is always possible to define a rotational matrix $\mathbf{R}_i$, transforming the coordinate     
to a local one~\cite{del2004quantum} with the local $y$-axis along $\langle \mathbf{S}_i\rangle$. 
The local spin $\mathbf{S}_i'$ can still be mapped into the ``phase space'' using Eq.~\eqref{Map}, 
then the same method described in the ferromagnetic example can be applied without technical obstacles.


There have been many triangular lattice magnets with anisotropic interactions that are under active
experimental study. These include the popular rare-earth triangular lattice antiferromagnets~\cite{PhysRevB.94.035107,Liu_2018}, 
and many transition metal compounds such as FeI$_2$ and CoI$_2$~\cite{Bai_2021,PhysRevLett.127.267201,chen2023quadrupole,kim2023bond}, et al.
Besides, the form of the spin model only relies on the discrete symmetries and the MEC 
occurs naturally from the lattice vibration. Thus, our result can be well extended to many anisotropic magnets 
that are not limited to exchange couplings and triangular lattices. 
In principle, the topological classification can be generally used 
for classifying and searching magnetoelastic coupled systems 
with chiral excitations. Among those materials, the one with the
bond-dependent Kitaev-type interactions on the honeycomb lattices~\cite{TREBST20221,rau2014generic,kocsis2022magnetoelastic,kim2023bond}, 
which may support a zigzag or a stripy antiferromagnetic ordered state~\cite{chaloupka2013zigzag} 
as well as the well-known Kitaev spin liquid state~\cite{kitaev2006anyons,takagi2019concept}, 
can be of particular interest. Magnetoelastic coupling anisotropy has been recently found 
in the Kitaev material $\alpha$-RuCl$_3$~\cite{kocsis2022magnetoelastic}, 
whereas the generalization of our model with fractionalized excitations \cite{zhang2021phonon} may need more careful considerations.

In summary, we have proposed the chiral magnon-phonon hybridizations induced from the bond-dependent 
magnetoelastic couplings by a general consideration of the lattice symmetries and 
provided a systematic way to understand this excitation by mapping the Hamiltonian into 
its phononic analog. In contrast to previous studies on chiral phonons with Raman spin-lattice coupling or topological hybridizations with magnetoelastic couplings, 
our mechanism does not require any out-of-plane magnetization or external field and thus could explain the planar thermal Hall results on CoTiO$_3$. Our study provides a new path for tuning chiral excitations with non-trivial topology 
in varieties of anisotropic-exchange materials and suggests possible applications 
in magneto-phononics and spin caloritronic.
\begin{acknowledgments}
\emph{Acknowledgment.}---We thank Yuxin Zhao for clarifying the definition of chirality, Gregory A. Fiete for discussions, and Jiaming He for sharing the thermal Hall data on CoTiO$_3$. This work is supported by the Collaborative Research Fund of Hong Kong with Grant No. C6009-20G and C7012-21G, by the Guangdong-Hong Kong Joint Laboratory of Quantum Matter, the NSFC/RGC JRS grant with Grant No. N$\_$HKU774/21, by the General Research Fund of Hong Kong with Grants No. 17310622 and No. 17303023, by the National Science Foundation of China with Grant No. 92065203, by the Ministry of Science and Technology of China with Grant No. 2021YFA1400300, and by the Fundamental Research Funds for the Central Universities, Peking University.
\end{acknowledgments}
\emph{Notes added.}---Upon preparing the manuscript, we become aware of a recent experimental work~\cite{cui2023chirality} observing the gap opening by the magnon-phonon interaction with a similar magnetoelastic coupling, while our model is a general one that does not require out-of-plane magnetic moments, and we theoretically point out the significance of the magnetoelastic coupling anisotropy on both chirality and topology of the excitations. 

\bibliography{CMP}

\providecommand{\noopsort}[1]{}\providecommand{\singleletter}[1]{#1}%
\begin{thebibliography}{86}%
\makeatletter
\providecommand \@ifxundefined [1]{%
 \@ifx{#1\undefined}
}%
\providecommand \@ifnum [1]{%
 \ifnum #1\expandafter \@firstoftwo
 \else \expandafter \@secondoftwo
 \fi
}%
\providecommand \@ifx [1]{%
 \ifx #1\expandafter \@firstoftwo
 \else \expandafter \@secondoftwo
 \fi
}%
\providecommand \natexlab [1]{#1}%
\providecommand \enquote  [1]{``#1''}%
\providecommand \bibnamefont  [1]{#1}%
\providecommand \bibfnamefont [1]{#1}%
\providecommand \citenamefont [1]{#1}%
\providecommand \href@noop [0]{\@secondoftwo}%
\providecommand \href [0]{\begingroup \@sanitize@url \@href}%
\providecommand \@href[1]{\@@startlink{#1}\@@href}%
\providecommand \@@href[1]{\endgroup#1\@@endlink}%
\providecommand \@sanitize@url [0]{\catcode `\\12\catcode `\$12\catcode `\&12\catcode `\#12\catcode `\^12\catcode `\_12\catcode `\%12\relax}%
\providecommand \@@startlink[1]{}%
\providecommand \@@endlink[0]{}%
\providecommand \url  [0]{\begingroup\@sanitize@url \@url }%
\providecommand \@url [1]{\endgroup\@href {#1}{\urlprefix }}%
\providecommand \urlprefix  [0]{URL }%
\providecommand \Eprint [0]{\href }%
\providecommand \doibase [0]{https://doi.org/}%
\providecommand \selectlanguage [0]{\@gobble}%
\providecommand \bibinfo  [0]{\@secondoftwo}%
\providecommand \bibfield  [0]{\@secondoftwo}%
\providecommand \translation [1]{[#1]}%
\providecommand \BibitemOpen [0]{}%
\providecommand \bibitemStop [0]{}%
\providecommand \bibitemNoStop [0]{.\EOS\space}%
\providecommand \EOS [0]{\spacefactor3000\relax}%
\providecommand \BibitemShut  [1]{\csname bibitem#1\endcsname}%
\let\auto@bib@innerbib\@empty
\bibitem [{\citenamefont {Zhang}\ and\ \citenamefont {Niu}(2015)}]{zhang2015chiral}%
  \BibitemOpen
  \bibfield  {author} {\bibinfo {author} {\bibfnamefont {L.}~\bibnamefont {Zhang}}\ and\ \bibinfo {author} {\bibfnamefont {Q.}~\bibnamefont {Niu}},\ }\bibfield  {title} {\bibinfo {title} {Chiral phonons at high-symmetry points in monolayer hexagonal lattices},\ }\href {https://doi.org/10.1103/PhysRevLett.115.115502} {\bibfield  {journal} {\bibinfo  {journal} {Phys. Rev. Lett.}\ }\textbf {\bibinfo {volume} {115}},\ \bibinfo {pages} {115502} (\bibinfo {year} {2015})}\BibitemShut {NoStop}%
\bibitem [{\citenamefont {Chen}\ \emph {et~al.}(2018)\citenamefont {Chen}, \citenamefont {Zhang}, \citenamefont {Niu},\ and\ \citenamefont {Zhang}}]{chen2018chiral}%
  \BibitemOpen
  \bibfield  {author} {\bibinfo {author} {\bibfnamefont {H.}~\bibnamefont {Chen}}, \bibinfo {author} {\bibfnamefont {W.}~\bibnamefont {Zhang}}, \bibinfo {author} {\bibfnamefont {Q.}~\bibnamefont {Niu}},\ and\ \bibinfo {author} {\bibfnamefont {L.}~\bibnamefont {Zhang}},\ }\bibfield  {title} {\bibinfo {title} {Chiral phonons in two-dimensional materials},\ }\href {https://doi.org/10.1088/2053-1583/aaf292} {\bibfield  {journal} {\bibinfo  {journal} {2D Materials}\ }\textbf {\bibinfo {volume} {6}},\ \bibinfo {pages} {012002} (\bibinfo {year} {2018})}\BibitemShut {NoStop}%
\bibitem [{\citenamefont {Gao}\ \emph {et~al.}(2018)\citenamefont {Gao}, \citenamefont {Zhang},\ and\ \citenamefont {Zhang}}]{gao2018nondegenerate}%
  \BibitemOpen
  \bibfield  {author} {\bibinfo {author} {\bibfnamefont {M.}~\bibnamefont {Gao}}, \bibinfo {author} {\bibfnamefont {W.}~\bibnamefont {Zhang}},\ and\ \bibinfo {author} {\bibfnamefont {L.}~\bibnamefont {Zhang}},\ }\bibfield  {title} {\bibinfo {title} {Nondegenerate chiral phonons in graphene/hexagonal boron nitride heterostructure from first-principles calculations},\ }\href {https://doi.org/10.1021/acs.nanolett.8b01487} {\bibfield  {journal} {\bibinfo  {journal} {Nano letters}\ }\textbf {\bibinfo {volume} {18}},\ \bibinfo {pages} {4424} (\bibinfo {year} {2018})}\BibitemShut {NoStop}%
\bibitem [{\citenamefont {Chen}\ \emph {et~al.}(2019{\natexlab{a}})\citenamefont {Chen}, \citenamefont {Wu}, \citenamefont {Yang}, \citenamefont {Li},\ and\ \citenamefont {Zhang}}]{chen2019chiral}%
  \BibitemOpen
  \bibfield  {author} {\bibinfo {author} {\bibfnamefont {H.}~\bibnamefont {Chen}}, \bibinfo {author} {\bibfnamefont {W.}~\bibnamefont {Wu}}, \bibinfo {author} {\bibfnamefont {S.~A.}\ \bibnamefont {Yang}}, \bibinfo {author} {\bibfnamefont {X.}~\bibnamefont {Li}},\ and\ \bibinfo {author} {\bibfnamefont {L.}~\bibnamefont {Zhang}},\ }\bibfield  {title} {\bibinfo {title} {Chiral phonons in kagome lattices},\ }\href {https://doi.org/10.1103/PhysRevB.100.094303} {\bibfield  {journal} {\bibinfo  {journal} {Phys. Rev. B}\ }\textbf {\bibinfo {volume} {100}},\ \bibinfo {pages} {094303} (\bibinfo {year} {2019}{\natexlab{a}})}\BibitemShut {NoStop}%
\bibitem [{\citenamefont {Saparov}\ \emph {et~al.}(2022)\citenamefont {Saparov}, \citenamefont {Xiong}, \citenamefont {Ren},\ and\ \citenamefont {Niu}}]{saparov2022lattice}%
  \BibitemOpen
  \bibfield  {author} {\bibinfo {author} {\bibfnamefont {D.}~\bibnamefont {Saparov}}, \bibinfo {author} {\bibfnamefont {B.}~\bibnamefont {Xiong}}, \bibinfo {author} {\bibfnamefont {Y.}~\bibnamefont {Ren}},\ and\ \bibinfo {author} {\bibfnamefont {Q.}~\bibnamefont {Niu}},\ }\bibfield  {title} {\bibinfo {title} {Lattice dynamics with molecular berry curvature: Chiral optical phonons},\ }\href {https://doi.org/10.1103/PhysRevB.105.064303} {\bibfield  {journal} {\bibinfo  {journal} {Phys. Rev. B}\ }\textbf {\bibinfo {volume} {105}},\ \bibinfo {pages} {064303} (\bibinfo {year} {2022})}\BibitemShut {NoStop}%
\bibitem [{\citenamefont {Zhang}\ and\ \citenamefont {Murakami}(2022)}]{zhang2022chiral}%
  \BibitemOpen
  \bibfield  {author} {\bibinfo {author} {\bibfnamefont {T.}~\bibnamefont {Zhang}}\ and\ \bibinfo {author} {\bibfnamefont {S.}~\bibnamefont {Murakami}},\ }\bibfield  {title} {\bibinfo {title} {Chiral phonons and pseudoangular momentum in nonsymmorphic systems},\ }\href {https://doi.org/10.1103/PhysRevResearch.4.L012024} {\bibfield  {journal} {\bibinfo  {journal} {Phys. Rev. Res.}\ }\textbf {\bibinfo {volume} {4}},\ \bibinfo {pages} {L012024} (\bibinfo {year} {2022})}\BibitemShut {NoStop}%
\bibitem [{\citenamefont {Zhu}\ \emph {et~al.}(2018)\citenamefont {Zhu}, \citenamefont {Yi}, \citenamefont {Li}, \citenamefont {Xiao}, \citenamefont {Zhang}, \citenamefont {Yang}, \citenamefont {Kaindl}, \citenamefont {Li}, \citenamefont {Wang},\ and\ \citenamefont {Zhang}}]{zhu2018observation}%
  \BibitemOpen
  \bibfield  {author} {\bibinfo {author} {\bibfnamefont {H.}~\bibnamefont {Zhu}}, \bibinfo {author} {\bibfnamefont {J.}~\bibnamefont {Yi}}, \bibinfo {author} {\bibfnamefont {M.-Y.}\ \bibnamefont {Li}}, \bibinfo {author} {\bibfnamefont {J.}~\bibnamefont {Xiao}}, \bibinfo {author} {\bibfnamefont {L.}~\bibnamefont {Zhang}}, \bibinfo {author} {\bibfnamefont {C.-W.}\ \bibnamefont {Yang}}, \bibinfo {author} {\bibfnamefont {R.~A.}\ \bibnamefont {Kaindl}}, \bibinfo {author} {\bibfnamefont {L.-J.}\ \bibnamefont {Li}}, \bibinfo {author} {\bibfnamefont {Y.}~\bibnamefont {Wang}},\ and\ \bibinfo {author} {\bibfnamefont {X.}~\bibnamefont {Zhang}},\ }\bibfield  {title} {\bibinfo {title} {Observation of chiral phonons},\ }\href {https://doi.org/10.1126/science.aar2711} {\bibfield  {journal} {\bibinfo  {journal} {Science}\ }\textbf {\bibinfo {volume} {359}},\ \bibinfo {pages} {579} (\bibinfo {year} {2018})}\BibitemShut {NoStop}%
\bibitem [{\citenamefont {Nakane}\ and\ \citenamefont {Kohno}(2018)}]{nakane2018angular}%
  \BibitemOpen
  \bibfield  {author} {\bibinfo {author} {\bibfnamefont {J.~J.}\ \bibnamefont {Nakane}}\ and\ \bibinfo {author} {\bibfnamefont {H.}~\bibnamefont {Kohno}},\ }\bibfield  {title} {\bibinfo {title} {Angular momentum of phonons and its application to single-spin relaxation},\ }\href {https://doi.org/10.1103/PhysRevB.97.174403} {\bibfield  {journal} {\bibinfo  {journal} {Phys. Rev. B}\ }\textbf {\bibinfo {volume} {97}},\ \bibinfo {pages} {174403} (\bibinfo {year} {2018})}\BibitemShut {NoStop}%
\bibitem [{\citenamefont {Chen}\ \emph {et~al.}(2019{\natexlab{b}})\citenamefont {Chen}, \citenamefont {Lu}, \citenamefont {Dubey}, \citenamefont {Yao}, \citenamefont {Liu}, \citenamefont {Wang}, \citenamefont {Xiong}, \citenamefont {Zhang},\ and\ \citenamefont {Srivastava}}]{chen2019entanglement}%
  \BibitemOpen
  \bibfield  {author} {\bibinfo {author} {\bibfnamefont {X.}~\bibnamefont {Chen}}, \bibinfo {author} {\bibfnamefont {X.}~\bibnamefont {Lu}}, \bibinfo {author} {\bibfnamefont {S.}~\bibnamefont {Dubey}}, \bibinfo {author} {\bibfnamefont {Q.}~\bibnamefont {Yao}}, \bibinfo {author} {\bibfnamefont {S.}~\bibnamefont {Liu}}, \bibinfo {author} {\bibfnamefont {X.}~\bibnamefont {Wang}}, \bibinfo {author} {\bibfnamefont {Q.}~\bibnamefont {Xiong}}, \bibinfo {author} {\bibfnamefont {L.}~\bibnamefont {Zhang}},\ and\ \bibinfo {author} {\bibfnamefont {A.}~\bibnamefont {Srivastava}},\ }\bibfield  {title} {\bibinfo {title} {Entanglement of single-photons and chiral phonons in atomically thin wse2},\ }\href {https://doi.org/10.1038/s41567-018-0366-7} {\bibfield  {journal} {\bibinfo  {journal} {Nature Physics}\ }\textbf {\bibinfo {volume} {15}},\ \bibinfo {pages} {221} (\bibinfo {year} {2019}{\natexlab{b}})}\BibitemShut {NoStop}%
\bibitem [{\citenamefont {Baydin}\ \emph {et~al.}(2022)\citenamefont {Baydin}, \citenamefont {Hernandez}, \citenamefont {Rodriguez-Vega}, \citenamefont {Okazaki}, \citenamefont {Tay}, \citenamefont {Noe}, \citenamefont {Katayama}, \citenamefont {Takeda}, \citenamefont {Nojiri}, \citenamefont {Rappl}, \citenamefont {Abramof}, \citenamefont {Fiete},\ and\ \citenamefont {Kono}}]{baydin2022magnetic}%
  \BibitemOpen
  \bibfield  {author} {\bibinfo {author} {\bibfnamefont {A.}~\bibnamefont {Baydin}}, \bibinfo {author} {\bibfnamefont {F.~G.~G.}\ \bibnamefont {Hernandez}}, \bibinfo {author} {\bibfnamefont {M.}~\bibnamefont {Rodriguez-Vega}}, \bibinfo {author} {\bibfnamefont {A.~K.}\ \bibnamefont {Okazaki}}, \bibinfo {author} {\bibfnamefont {F.}~\bibnamefont {Tay}}, \bibinfo {author} {\bibfnamefont {G.~T.}\ \bibnamefont {Noe}}, \bibinfo {author} {\bibfnamefont {I.}~\bibnamefont {Katayama}}, \bibinfo {author} {\bibfnamefont {J.}~\bibnamefont {Takeda}}, \bibinfo {author} {\bibfnamefont {H.}~\bibnamefont {Nojiri}}, \bibinfo {author} {\bibfnamefont {P.~H.~O.}\ \bibnamefont {Rappl}}, \bibinfo {author} {\bibfnamefont {E.}~\bibnamefont {Abramof}}, \bibinfo {author} {\bibfnamefont {G.~A.}\ \bibnamefont {Fiete}},\ and\ \bibinfo {author} {\bibfnamefont {J.}~\bibnamefont {Kono}},\ }\bibfield  {title} {\bibinfo {title} {Magnetic control of soft chiral phonons in pbte},\ }\href {https://doi.org/10.1103/PhysRevLett.128.075901} {\bibfield
  {journal} {\bibinfo  {journal} {Phys. Rev. Lett.}\ }\textbf {\bibinfo {volume} {128}},\ \bibinfo {pages} {075901} (\bibinfo {year} {2022})}\BibitemShut {NoStop}%
\bibitem [{\citenamefont {Ishito}\ \emph {et~al.}(2023)\citenamefont {Ishito}, \citenamefont {Mao}, \citenamefont {Kousaka}, \citenamefont {Togawa}, \citenamefont {Iwasaki}, \citenamefont {Zhang}, \citenamefont {Murakami}, \citenamefont {Kishine},\ and\ \citenamefont {Satoh}}]{ishito2023truly}%
  \BibitemOpen
  \bibfield  {author} {\bibinfo {author} {\bibfnamefont {K.}~\bibnamefont {Ishito}}, \bibinfo {author} {\bibfnamefont {H.}~\bibnamefont {Mao}}, \bibinfo {author} {\bibfnamefont {Y.}~\bibnamefont {Kousaka}}, \bibinfo {author} {\bibfnamefont {Y.}~\bibnamefont {Togawa}}, \bibinfo {author} {\bibfnamefont {S.}~\bibnamefont {Iwasaki}}, \bibinfo {author} {\bibfnamefont {T.}~\bibnamefont {Zhang}}, \bibinfo {author} {\bibfnamefont {S.}~\bibnamefont {Murakami}}, \bibinfo {author} {\bibfnamefont {J.-i.}\ \bibnamefont {Kishine}},\ and\ \bibinfo {author} {\bibfnamefont {T.}~\bibnamefont {Satoh}},\ }\bibfield  {title} {\bibinfo {title} {Truly chiral phonons in $\alpha$-hgs},\ }\href {https://doi.org/10.1038/s41567-022-01790-x} {\bibfield  {journal} {\bibinfo  {journal} {Nature Physics}\ }\textbf {\bibinfo {volume} {19}},\ \bibinfo {pages} {35} (\bibinfo {year} {2023})}\BibitemShut {NoStop}%
\bibitem [{\citenamefont {Ueda}\ \emph {et~al.}(2023)\citenamefont {Ueda}, \citenamefont {Garc{\'\i}a-Fern{\'a}ndez}, \citenamefont {Agrestini}, \citenamefont {Romao}, \citenamefont {van~den Brink}, \citenamefont {Spaldin}, \citenamefont {Zhou},\ and\ \citenamefont {Staub}}]{ueda2023chiral}%
  \BibitemOpen
  \bibfield  {author} {\bibinfo {author} {\bibfnamefont {H.}~\bibnamefont {Ueda}}, \bibinfo {author} {\bibfnamefont {M.}~\bibnamefont {Garc{\'\i}a-Fern{\'a}ndez}}, \bibinfo {author} {\bibfnamefont {S.}~\bibnamefont {Agrestini}}, \bibinfo {author} {\bibfnamefont {C.~P.}\ \bibnamefont {Romao}}, \bibinfo {author} {\bibfnamefont {J.}~\bibnamefont {van~den Brink}}, \bibinfo {author} {\bibfnamefont {N.~A.}\ \bibnamefont {Spaldin}}, \bibinfo {author} {\bibfnamefont {K.-J.}\ \bibnamefont {Zhou}},\ and\ \bibinfo {author} {\bibfnamefont {U.}~\bibnamefont {Staub}},\ }\bibfield  {title} {\bibinfo {title} {Chiral phonons in quartz probed by x-rays},\ }\href {https://doi.org/10.1038/s41586-023-06016-5} {\bibfield  {journal} {\bibinfo  {journal} {Nature}\ ,\ \bibinfo {pages} {1}} (\bibinfo {year} {2023})}\BibitemShut {NoStop}%
\bibitem [{\citenamefont {Zhang}\ and\ \citenamefont {Niu}(2014)}]{zhang2014angular}%
  \BibitemOpen
  \bibfield  {author} {\bibinfo {author} {\bibfnamefont {L.}~\bibnamefont {Zhang}}\ and\ \bibinfo {author} {\bibfnamefont {Q.}~\bibnamefont {Niu}},\ }\bibfield  {title} {\bibinfo {title} {Angular momentum of phonons and the einstein--de haas effect},\ }\href {https://doi.org/10.1103/PhysRevLett.112.085503} {\bibfield  {journal} {\bibinfo  {journal} {Phys. Rev. Lett.}\ }\textbf {\bibinfo {volume} {112}},\ \bibinfo {pages} {085503} (\bibinfo {year} {2014})}\BibitemShut {NoStop}%
\bibitem [{\citenamefont {Mentink}\ \emph {et~al.}(2019)\citenamefont {Mentink}, \citenamefont {Katsnelson},\ and\ \citenamefont {Lemeshko}}]{mentink2019quantum}%
  \BibitemOpen
  \bibfield  {author} {\bibinfo {author} {\bibfnamefont {J.~H.}\ \bibnamefont {Mentink}}, \bibinfo {author} {\bibfnamefont {M.~I.}\ \bibnamefont {Katsnelson}},\ and\ \bibinfo {author} {\bibfnamefont {M.}~\bibnamefont {Lemeshko}},\ }\bibfield  {title} {\bibinfo {title} {Quantum many-body dynamics of the einstein--de haas effect},\ }\href {https://doi.org/10.1103/PhysRevB.99.064428} {\bibfield  {journal} {\bibinfo  {journal} {Phys. Rev. B}\ }\textbf {\bibinfo {volume} {99}},\ \bibinfo {pages} {064428} (\bibinfo {year} {2019})}\BibitemShut {NoStop}%
\bibitem [{\citenamefont {Dornes}\ \emph {et~al.}(2019)\citenamefont {Dornes}, \citenamefont {Acremann}, \citenamefont {Savoini}, \citenamefont {Kubli}, \citenamefont {Neugebauer}, \citenamefont {Abreu}, \citenamefont {Huber}, \citenamefont {Lantz}, \citenamefont {Vaz}, \citenamefont {Lemke} \emph {et~al.}}]{dornes2019ultrafast}%
  \BibitemOpen
  \bibfield  {author} {\bibinfo {author} {\bibfnamefont {C.}~\bibnamefont {Dornes}}, \bibinfo {author} {\bibfnamefont {Y.}~\bibnamefont {Acremann}}, \bibinfo {author} {\bibfnamefont {M.}~\bibnamefont {Savoini}}, \bibinfo {author} {\bibfnamefont {M.}~\bibnamefont {Kubli}}, \bibinfo {author} {\bibfnamefont {M.~J.}\ \bibnamefont {Neugebauer}}, \bibinfo {author} {\bibfnamefont {E.}~\bibnamefont {Abreu}}, \bibinfo {author} {\bibfnamefont {L.}~\bibnamefont {Huber}}, \bibinfo {author} {\bibfnamefont {G.}~\bibnamefont {Lantz}}, \bibinfo {author} {\bibfnamefont {C.~A.}\ \bibnamefont {Vaz}}, \bibinfo {author} {\bibfnamefont {H.}~\bibnamefont {Lemke}}, \emph {et~al.},\ }\bibfield  {title} {\bibinfo {title} {The ultrafast einstein--de haas effect},\ }\href {https://doi.org/10.1038/s41586-018-0822-7} {\bibfield  {journal} {\bibinfo  {journal} {Nature}\ }\textbf {\bibinfo {volume} {565}},\ \bibinfo {pages} {209} (\bibinfo {year} {2019})}\BibitemShut {NoStop}%
\bibitem [{\citenamefont {Ren}\ \emph {et~al.}(2021)\citenamefont {Ren}, \citenamefont {Xiao}, \citenamefont {Saparov},\ and\ \citenamefont {Niu}}]{ren2021phonon}%
  \BibitemOpen
  \bibfield  {author} {\bibinfo {author} {\bibfnamefont {Y.}~\bibnamefont {Ren}}, \bibinfo {author} {\bibfnamefont {C.}~\bibnamefont {Xiao}}, \bibinfo {author} {\bibfnamefont {D.}~\bibnamefont {Saparov}},\ and\ \bibinfo {author} {\bibfnamefont {Q.}~\bibnamefont {Niu}},\ }\bibfield  {title} {\bibinfo {title} {Phonon magnetic moment from electronic topological magnetization},\ }\href {https://doi.org/10.1103/PhysRevLett.127.186403} {\bibfield  {journal} {\bibinfo  {journal} {Phys. Rev. Lett.}\ }\textbf {\bibinfo {volume} {127}},\ \bibinfo {pages} {186403} (\bibinfo {year} {2021})}\BibitemShut {NoStop}%
\bibitem [{\citenamefont {Xiong}\ \emph {et~al.}(2022)\citenamefont {Xiong}, \citenamefont {Chen}, \citenamefont {Ma},\ and\ \citenamefont {Zhang}}]{xiong2022effective}%
  \BibitemOpen
  \bibfield  {author} {\bibinfo {author} {\bibfnamefont {G.}~\bibnamefont {Xiong}}, \bibinfo {author} {\bibfnamefont {H.}~\bibnamefont {Chen}}, \bibinfo {author} {\bibfnamefont {D.}~\bibnamefont {Ma}},\ and\ \bibinfo {author} {\bibfnamefont {L.}~\bibnamefont {Zhang}},\ }\bibfield  {title} {\bibinfo {title} {Effective magnetic fields induced by chiral phonons},\ }\href {https://doi.org/10.1103/PhysRevB.106.144302} {\bibfield  {journal} {\bibinfo  {journal} {Phys. Rev. B}\ }\textbf {\bibinfo {volume} {106}},\ \bibinfo {pages} {144302} (\bibinfo {year} {2022})}\BibitemShut {NoStop}%
\bibitem [{\citenamefont {Fransson}(2023)}]{fransson2023chiral}%
  \BibitemOpen
  \bibfield  {author} {\bibinfo {author} {\bibfnamefont {J.}~\bibnamefont {Fransson}},\ }\bibfield  {title} {\bibinfo {title} {Chiral phonon induced spin polarization},\ }\href {https://doi.org/10.1103/PhysRevResearch.5.L022039} {\bibfield  {journal} {\bibinfo  {journal} {Phys. Rev. Res.}\ }\textbf {\bibinfo {volume} {5}},\ \bibinfo {pages} {L022039} (\bibinfo {year} {2023})}\BibitemShut {NoStop}%
\bibitem [{\citenamefont {Hernandez}\ \emph {et~al.}(2023)\citenamefont {Hernandez}, \citenamefont {Baydin}, \citenamefont {Chaudhary}, \citenamefont {Tay}, \citenamefont {Katayama}, \citenamefont {Takeda}, \citenamefont {Nojiri}, \citenamefont {Okazaki}, \citenamefont {Rappl}, \citenamefont {Abramof} \emph {et~al.}}]{hernandez2023observation}%
  \BibitemOpen
  \bibfield  {author} {\bibinfo {author} {\bibfnamefont {F.~G.}\ \bibnamefont {Hernandez}}, \bibinfo {author} {\bibfnamefont {A.}~\bibnamefont {Baydin}}, \bibinfo {author} {\bibfnamefont {S.}~\bibnamefont {Chaudhary}}, \bibinfo {author} {\bibfnamefont {F.}~\bibnamefont {Tay}}, \bibinfo {author} {\bibfnamefont {I.}~\bibnamefont {Katayama}}, \bibinfo {author} {\bibfnamefont {J.}~\bibnamefont {Takeda}}, \bibinfo {author} {\bibfnamefont {H.}~\bibnamefont {Nojiri}}, \bibinfo {author} {\bibfnamefont {A.~K.}\ \bibnamefont {Okazaki}}, \bibinfo {author} {\bibfnamefont {P.~H.}\ \bibnamefont {Rappl}}, \bibinfo {author} {\bibfnamefont {E.}~\bibnamefont {Abramof}}, \emph {et~al.},\ }\bibfield  {title} {\bibinfo {title} {Observation of interplay between phonon chirality and electronic band topology},\ }\href {https://doi.org/10.1126/sciadv.adj4074} {\bibfield  {journal} {\bibinfo  {journal} {Science advances}\ }\textbf {\bibinfo {volume} {9}},\ \bibinfo {pages} {eadj4074} (\bibinfo {year} {2023})}\BibitemShut {NoStop}%
\bibitem [{\citenamefont {Chaudhary}\ \emph {et~al.}(2023)\citenamefont {Chaudhary}, \citenamefont {Juraschek}, \citenamefont {Rodriguez-Vega},\ and\ \citenamefont {Fiete}}]{chaudhary2023giant}%
  \BibitemOpen
  \bibfield  {author} {\bibinfo {author} {\bibfnamefont {S.}~\bibnamefont {Chaudhary}}, \bibinfo {author} {\bibfnamefont {D.~M.}\ \bibnamefont {Juraschek}}, \bibinfo {author} {\bibfnamefont {M.}~\bibnamefont {Rodriguez-Vega}},\ and\ \bibinfo {author} {\bibfnamefont {G.~A.}\ \bibnamefont {Fiete}},\ }\bibfield  {title} {\bibinfo {title} {Giant effective magnetic moments of chiral phonons from orbit-lattice coupling},\ }\href {https://arxiv.org/abs/2306.11630} {\bibfield  {journal} {\bibinfo  {journal} {arXiv preprint arXiv:2306.11630}\ } (\bibinfo {year} {2023})}\BibitemShut {NoStop}%
\bibitem [{\citenamefont {Wu}\ \emph {et~al.}(2023)\citenamefont {Wu}, \citenamefont {Bao}, \citenamefont {Zhou}, \citenamefont {Wang}, \citenamefont {Sun}, \citenamefont {Wen}, \citenamefont {Wan},\ and\ \citenamefont {Zhang}}]{wu2023fluctuation}%
  \BibitemOpen
  \bibfield  {author} {\bibinfo {author} {\bibfnamefont {F.}~\bibnamefont {Wu}}, \bibinfo {author} {\bibfnamefont {S.}~\bibnamefont {Bao}}, \bibinfo {author} {\bibfnamefont {J.}~\bibnamefont {Zhou}}, \bibinfo {author} {\bibfnamefont {Y.}~\bibnamefont {Wang}}, \bibinfo {author} {\bibfnamefont {J.}~\bibnamefont {Sun}}, \bibinfo {author} {\bibfnamefont {J.}~\bibnamefont {Wen}}, \bibinfo {author} {\bibfnamefont {Y.}~\bibnamefont {Wan}},\ and\ \bibinfo {author} {\bibfnamefont {Q.}~\bibnamefont {Zhang}},\ }\bibfield  {title} {\bibinfo {title} {Fluctuation-enhanced phonon magnetic moments in a polar antiferromagnet},\ }\href {https://doi.org/10.1038/s41567-023-02210-4} {\bibfield  {journal} {\bibinfo  {journal} {Nature Physics}\ ,\ \bibinfo {pages} {1}} (\bibinfo {year} {2023})}\BibitemShut {NoStop}%
\bibitem [{\citenamefont {Juraschek}\ and\ \citenamefont {Spaldin}(2019)}]{juraschek2019orbital}%
  \BibitemOpen
  \bibfield  {author} {\bibinfo {author} {\bibfnamefont {D.~M.}\ \bibnamefont {Juraschek}}\ and\ \bibinfo {author} {\bibfnamefont {N.~A.}\ \bibnamefont {Spaldin}},\ }\bibfield  {title} {\bibinfo {title} {Orbital magnetic moments of phonons},\ }\href {https://doi.org/10.1103/PhysRevMaterials.3.064405} {\bibfield  {journal} {\bibinfo  {journal} {Phys. Rev. Mater.}\ }\textbf {\bibinfo {volume} {3}},\ \bibinfo {pages} {064405} (\bibinfo {year} {2019})}\BibitemShut {NoStop}%
\bibitem [{\citenamefont {Juraschek}\ \emph {et~al.}(2020)\citenamefont {Juraschek}, \citenamefont {Narang},\ and\ \citenamefont {Spaldin}}]{juraschek2020phono}%
  \BibitemOpen
  \bibfield  {author} {\bibinfo {author} {\bibfnamefont {D.~M.}\ \bibnamefont {Juraschek}}, \bibinfo {author} {\bibfnamefont {P.}~\bibnamefont {Narang}},\ and\ \bibinfo {author} {\bibfnamefont {N.~A.}\ \bibnamefont {Spaldin}},\ }\bibfield  {title} {\bibinfo {title} {Phono-magnetic analogs to opto-magnetic effects},\ }\href {https://doi.org/10.1103/PhysRevResearch.2.043035} {\bibfield  {journal} {\bibinfo  {journal} {Phys. Rev. Res.}\ }\textbf {\bibinfo {volume} {2}},\ \bibinfo {pages} {043035} (\bibinfo {year} {2020})}\BibitemShut {NoStop}%
\bibitem [{\citenamefont {Juraschek}\ \emph {et~al.}(2022)\citenamefont {Juraschek}, \citenamefont {Neuman},\ and\ \citenamefont {Narang}}]{juraschek2022giant}%
  \BibitemOpen
  \bibfield  {author} {\bibinfo {author} {\bibfnamefont {D.~M.}\ \bibnamefont {Juraschek}}, \bibinfo {author} {\bibfnamefont {T.~c.~v.}\ \bibnamefont {Neuman}},\ and\ \bibinfo {author} {\bibfnamefont {P.}~\bibnamefont {Narang}},\ }\bibfield  {title} {\bibinfo {title} {Giant effective magnetic fields from optically driven chiral phonons in $4f$ paramagnets},\ }\href {https://doi.org/10.1103/PhysRevResearch.4.013129} {\bibfield  {journal} {\bibinfo  {journal} {Phys. Rev. Res.}\ }\textbf {\bibinfo {volume} {4}},\ \bibinfo {pages} {013129} (\bibinfo {year} {2022})}\BibitemShut {NoStop}%
\bibitem [{\citenamefont {Wang}\ and\ \citenamefont {Li}(2008{\natexlab{a}})}]{wang2008phononics}%
  \BibitemOpen
  \bibfield  {author} {\bibinfo {author} {\bibfnamefont {L.}~\bibnamefont {Wang}}\ and\ \bibinfo {author} {\bibfnamefont {B.}~\bibnamefont {Li}},\ }\bibfield  {title} {\bibinfo {title} {Phononics gets hot},\ }\href {https://doi.org/10.1088/2058-7058/21/03/31} {\bibfield  {journal} {\bibinfo  {journal} {Physics World}\ }\textbf {\bibinfo {volume} {21}},\ \bibinfo {pages} {27} (\bibinfo {year} {2008}{\natexlab{a}})}\BibitemShut {NoStop}%
\bibitem [{\citenamefont {Wang}\ and\ \citenamefont {Li}(2008{\natexlab{b}})}]{wang2008thermal}%
  \BibitemOpen
  \bibfield  {author} {\bibinfo {author} {\bibfnamefont {L.}~\bibnamefont {Wang}}\ and\ \bibinfo {author} {\bibfnamefont {B.}~\bibnamefont {Li}},\ }\bibfield  {title} {\bibinfo {title} {Thermal memory: A storage of phononic information},\ }\href {https://doi.org/10.1103/PhysRevLett.101.267203} {\bibfield  {journal} {\bibinfo  {journal} {Phys. Rev. Lett.}\ }\textbf {\bibinfo {volume} {101}},\ \bibinfo {pages} {267203} (\bibinfo {year} {2008}{\natexlab{b}})}\BibitemShut {NoStop}%
\bibitem [{\citenamefont {Sheng}\ \emph {et~al.}(2006)\citenamefont {Sheng}, \citenamefont {Sheng},\ and\ \citenamefont {Ting}}]{sheng2006theory}%
  \BibitemOpen
  \bibfield  {author} {\bibinfo {author} {\bibfnamefont {L.}~\bibnamefont {Sheng}}, \bibinfo {author} {\bibfnamefont {D.~N.}\ \bibnamefont {Sheng}},\ and\ \bibinfo {author} {\bibfnamefont {C.~S.}\ \bibnamefont {Ting}},\ }\bibfield  {title} {\bibinfo {title} {Theory of the phonon hall effect in paramagnetic dielectrics},\ }\href {https://doi.org/10.1103/PhysRevLett.96.155901} {\bibfield  {journal} {\bibinfo  {journal} {Phys. Rev. Lett.}\ }\textbf {\bibinfo {volume} {96}},\ \bibinfo {pages} {155901} (\bibinfo {year} {2006})}\BibitemShut {NoStop}%
\bibitem [{\citenamefont {Ray}\ and\ \citenamefont {Ray}(1967)}]{ray1967dynamical}%
  \BibitemOpen
  \bibfield  {author} {\bibinfo {author} {\bibfnamefont {T.}~\bibnamefont {Ray}}\ and\ \bibinfo {author} {\bibfnamefont {D.~K.}\ \bibnamefont {Ray}},\ }\bibfield  {title} {\bibinfo {title} {Dynamical spin hamiltonian and the anisotropy of spin-lattice relaxation for the kramers doublets. i. general considerations},\ }\href {https://doi.org/10.1103/PhysRev.164.420} {\bibfield  {journal} {\bibinfo  {journal} {Phys. Rev.}\ }\textbf {\bibinfo {volume} {164}},\ \bibinfo {pages} {420} (\bibinfo {year} {1967})}\BibitemShut {NoStop}%
\bibitem [{\citenamefont {Ioselevich}\ and\ \citenamefont {Capellmann}(1995)}]{ioselevich1995strongly}%
  \BibitemOpen
  \bibfield  {author} {\bibinfo {author} {\bibfnamefont {A.~S.}\ \bibnamefont {Ioselevich}}\ and\ \bibinfo {author} {\bibfnamefont {H.}~\bibnamefont {Capellmann}},\ }\bibfield  {title} {\bibinfo {title} {Strongly correlated spin-phonon systems: A scenario for heavy fermions},\ }\href {https://doi.org/10.1103/PhysRevB.51.11446} {\bibfield  {journal} {\bibinfo  {journal} {Phys. Rev. B}\ }\textbf {\bibinfo {volume} {51}},\ \bibinfo {pages} {11446} (\bibinfo {year} {1995})}\BibitemShut {NoStop}%
\bibitem [{\citenamefont {Penc}\ \emph {et~al.}(2004)\citenamefont {Penc}, \citenamefont {Shannon},\ and\ \citenamefont {Shiba}}]{penc2004half}%
  \BibitemOpen
  \bibfield  {author} {\bibinfo {author} {\bibfnamefont {K.}~\bibnamefont {Penc}}, \bibinfo {author} {\bibfnamefont {N.}~\bibnamefont {Shannon}},\ and\ \bibinfo {author} {\bibfnamefont {H.}~\bibnamefont {Shiba}},\ }\bibfield  {title} {\bibinfo {title} {Half-magnetization plateau stabilized by structural distortion in the antiferromagnetic heisenberg model on a pyrochlore lattice},\ }\href {https://doi.org/10.1103/PhysRevLett.93.197203} {\bibfield  {journal} {\bibinfo  {journal} {Phys. Rev. Lett.}\ }\textbf {\bibinfo {volume} {93}},\ \bibinfo {pages} {197203} (\bibinfo {year} {2004})}\BibitemShut {NoStop}%
\bibitem [{JMH()}]{JMH}%
  \BibitemOpen
  \href@noop {} {}\bibinfo {note} {Private communication with Dr. Jiaming He}\BibitemShut {NoStop}%
\bibitem [{\citenamefont {Kishine}\ \emph {et~al.}(2020)\citenamefont {Kishine}, \citenamefont {Ovchinnikov},\ and\ \citenamefont {Tereshchenko}}]{kishine2020chirality}%
  \BibitemOpen
  \bibfield  {author} {\bibinfo {author} {\bibfnamefont {J.}~\bibnamefont {Kishine}}, \bibinfo {author} {\bibfnamefont {A.~S.}\ \bibnamefont {Ovchinnikov}},\ and\ \bibinfo {author} {\bibfnamefont {A.~A.}\ \bibnamefont {Tereshchenko}},\ }\bibfield  {title} {\bibinfo {title} {Chirality-induced phonon dispersion in a noncentrosymmetric micropolar crystal},\ }\href {https://doi.org/10.1103/PhysRevLett.125.245302} {\bibfield  {journal} {\bibinfo  {journal} {Phys. Rev. Lett.}\ }\textbf {\bibinfo {volume} {125}},\ \bibinfo {pages} {245302} (\bibinfo {year} {2020})}\BibitemShut {NoStop}%
\bibitem [{\citenamefont {Bonini}\ \emph {et~al.}(2023)\citenamefont {Bonini}, \citenamefont {Ren}, \citenamefont {Vanderbilt}, \citenamefont {Stengel}, \citenamefont {Dreyer},\ and\ \citenamefont {Coh}}]{bonini2023frequency}%
  \BibitemOpen
  \bibfield  {author} {\bibinfo {author} {\bibfnamefont {J.}~\bibnamefont {Bonini}}, \bibinfo {author} {\bibfnamefont {S.}~\bibnamefont {Ren}}, \bibinfo {author} {\bibfnamefont {D.}~\bibnamefont {Vanderbilt}}, \bibinfo {author} {\bibfnamefont {M.}~\bibnamefont {Stengel}}, \bibinfo {author} {\bibfnamefont {C.~E.}\ \bibnamefont {Dreyer}},\ and\ \bibinfo {author} {\bibfnamefont {S.}~\bibnamefont {Coh}},\ }\bibfield  {title} {\bibinfo {title} {Frequency splitting of chiral phonons from broken time-reversal symmetry in ${\mathrm{cri}}_{3}$},\ }\href {https://doi.org/10.1103/PhysRevLett.130.086701} {\bibfield  {journal} {\bibinfo  {journal} {Phys. Rev. Lett.}\ }\textbf {\bibinfo {volume} {130}},\ \bibinfo {pages} {086701} (\bibinfo {year} {2023})}\BibitemShut {NoStop}%
\bibitem [{\citenamefont {Takahashi}\ and\ \citenamefont {Nagaosa}(2016)}]{takahashi2016berry}%
  \BibitemOpen
  \bibfield  {author} {\bibinfo {author} {\bibfnamefont {R.}~\bibnamefont {Takahashi}}\ and\ \bibinfo {author} {\bibfnamefont {N.}~\bibnamefont {Nagaosa}},\ }\bibfield  {title} {\bibinfo {title} {Berry curvature in magnon-phonon hybrid systems},\ }\href {https://doi.org/10.1103/PhysRevLett.117.217205} {\bibfield  {journal} {\bibinfo  {journal} {Phys. Rev. Lett.}\ }\textbf {\bibinfo {volume} {117}},\ \bibinfo {pages} {217205} (\bibinfo {year} {2016})}\BibitemShut {NoStop}%
\bibitem [{\citenamefont {Zhang}\ \emph {et~al.}(2019)\citenamefont {Zhang}, \citenamefont {Zhang}, \citenamefont {Okamoto},\ and\ \citenamefont {Xiao}}]{zhang2019thermal}%
  \BibitemOpen
  \bibfield  {author} {\bibinfo {author} {\bibfnamefont {X.}~\bibnamefont {Zhang}}, \bibinfo {author} {\bibfnamefont {Y.}~\bibnamefont {Zhang}}, \bibinfo {author} {\bibfnamefont {S.}~\bibnamefont {Okamoto}},\ and\ \bibinfo {author} {\bibfnamefont {D.}~\bibnamefont {Xiao}},\ }\bibfield  {title} {\bibinfo {title} {Thermal hall effect induced by magnon-phonon interactions},\ }\href {https://doi.org/10.1103/PhysRevLett.123.167202} {\bibfield  {journal} {\bibinfo  {journal} {Phys. Rev. Lett.}\ }\textbf {\bibinfo {volume} {123}},\ \bibinfo {pages} {167202} (\bibinfo {year} {2019})}\BibitemShut {NoStop}%
\bibitem [{\citenamefont {Thingstad}\ \emph {et~al.}(2019)\citenamefont {Thingstad}, \citenamefont {Kamra}, \citenamefont {Brataas},\ and\ \citenamefont {Sudb\o{}}}]{thingstad2019chiral}%
  \BibitemOpen
  \bibfield  {author} {\bibinfo {author} {\bibfnamefont {E.}~\bibnamefont {Thingstad}}, \bibinfo {author} {\bibfnamefont {A.}~\bibnamefont {Kamra}}, \bibinfo {author} {\bibfnamefont {A.}~\bibnamefont {Brataas}},\ and\ \bibinfo {author} {\bibfnamefont {A.}~\bibnamefont {Sudb\o{}}},\ }\bibfield  {title} {\bibinfo {title} {Chiral phonon transport induced by topological magnons},\ }\href {https://doi.org/10.1103/PhysRevLett.122.107201} {\bibfield  {journal} {\bibinfo  {journal} {Phys. Rev. Lett.}\ }\textbf {\bibinfo {volume} {122}},\ \bibinfo {pages} {107201} (\bibinfo {year} {2019})}\BibitemShut {NoStop}%
\bibitem [{\citenamefont {Go}\ \emph {et~al.}(2019)\citenamefont {Go}, \citenamefont {Kim},\ and\ \citenamefont {Lee}}]{go2019topological}%
  \BibitemOpen
  \bibfield  {author} {\bibinfo {author} {\bibfnamefont {G.}~\bibnamefont {Go}}, \bibinfo {author} {\bibfnamefont {S.~K.}\ \bibnamefont {Kim}},\ and\ \bibinfo {author} {\bibfnamefont {K.-J.}\ \bibnamefont {Lee}},\ }\bibfield  {title} {\bibinfo {title} {Topological magnon-phonon hybrid excitations in two-dimensional ferromagnets with tunable chern numbers},\ }\href {https://doi.org/10.1103/PhysRevLett.123.237207} {\bibfield  {journal} {\bibinfo  {journal} {Phys. Rev. Lett.}\ }\textbf {\bibinfo {volume} {123}},\ \bibinfo {pages} {237207} (\bibinfo {year} {2019})}\BibitemShut {NoStop}%
\bibitem [{\citenamefont {Park}\ and\ \citenamefont {Yang}(2019)}]{park2019topological}%
  \BibitemOpen
  \bibfield  {author} {\bibinfo {author} {\bibfnamefont {S.}~\bibnamefont {Park}}\ and\ \bibinfo {author} {\bibfnamefont {B.-J.}\ \bibnamefont {Yang}},\ }\bibfield  {title} {\bibinfo {title} {Topological magnetoelastic excitations in noncollinear antiferromagnets},\ }\href {https://doi.org/10.1103/PhysRevB.99.174435} {\bibfield  {journal} {\bibinfo  {journal} {Phys. Rev. B}\ }\textbf {\bibinfo {volume} {99}},\ \bibinfo {pages} {174435} (\bibinfo {year} {2019})}\BibitemShut {NoStop}%
\bibitem [{\citenamefont {Sheikhi}\ \emph {et~al.}(2021)\citenamefont {Sheikhi}, \citenamefont {Kargarian},\ and\ \citenamefont {Langari}}]{sheikhi2021hybrid}%
  \BibitemOpen
  \bibfield  {author} {\bibinfo {author} {\bibfnamefont {B.}~\bibnamefont {Sheikhi}}, \bibinfo {author} {\bibfnamefont {M.}~\bibnamefont {Kargarian}},\ and\ \bibinfo {author} {\bibfnamefont {A.}~\bibnamefont {Langari}},\ }\bibfield  {title} {\bibinfo {title} {Hybrid topological magnon-phonon modes in ferromagnetic honeycomb and kagome lattices},\ }\href {https://doi.org/10.1103/PhysRevB.104.045139} {\bibfield  {journal} {\bibinfo  {journal} {Phys. Rev. B}\ }\textbf {\bibinfo {volume} {104}},\ \bibinfo {pages} {045139} (\bibinfo {year} {2021})}\BibitemShut {NoStop}%
\bibitem [{\citenamefont {Efimkin}\ and\ \citenamefont {Kargarian}(2021)}]{efimkin2021topological}%
  \BibitemOpen
  \bibfield  {author} {\bibinfo {author} {\bibfnamefont {D.~K.}\ \bibnamefont {Efimkin}}\ and\ \bibinfo {author} {\bibfnamefont {M.}~\bibnamefont {Kargarian}},\ }\bibfield  {title} {\bibinfo {title} {Topological spin-plasma waves},\ }\href {https://doi.org/10.1103/PhysRevB.104.075413} {\bibfield  {journal} {\bibinfo  {journal} {Phys. Rev. B}\ }\textbf {\bibinfo {volume} {104}},\ \bibinfo {pages} {075413} (\bibinfo {year} {2021})}\BibitemShut {NoStop}%
\bibitem [{\citenamefont {Ma}\ and\ \citenamefont {Fiete}(2022)}]{ma2022antiferromagnetic}%
  \BibitemOpen
  \bibfield  {author} {\bibinfo {author} {\bibfnamefont {B.}~\bibnamefont {Ma}}\ and\ \bibinfo {author} {\bibfnamefont {G.~A.}\ \bibnamefont {Fiete}},\ }\bibfield  {title} {\bibinfo {title} {Antiferromagnetic insulators with tunable magnon-polaron chern numbers induced by in-plane optical phonons},\ }\href {https://doi.org/10.1103/PhysRevB.105.L100402} {\bibfield  {journal} {\bibinfo  {journal} {Phys. Rev. B}\ }\textbf {\bibinfo {volume} {105}},\ \bibinfo {pages} {L100402} (\bibinfo {year} {2022})}\BibitemShut {NoStop}%
\bibitem [{\citenamefont {Bao}\ \emph {et~al.}(2020)\citenamefont {Bao}, \citenamefont {Cai}, \citenamefont {Si}, \citenamefont {Wang}, \citenamefont {Wang}, \citenamefont {Shangguan}, \citenamefont {Ma}, \citenamefont {Dong}, \citenamefont {Kajimoto}, \citenamefont {Ikeuchi}, \citenamefont {Yu}, \citenamefont {Sun}, \citenamefont {Li},\ and\ \citenamefont {Wen}}]{bao2020evidence}%
  \BibitemOpen
  \bibfield  {author} {\bibinfo {author} {\bibfnamefont {S.}~\bibnamefont {Bao}}, \bibinfo {author} {\bibfnamefont {Z.}~\bibnamefont {Cai}}, \bibinfo {author} {\bibfnamefont {W.}~\bibnamefont {Si}}, \bibinfo {author} {\bibfnamefont {W.}~\bibnamefont {Wang}}, \bibinfo {author} {\bibfnamefont {X.}~\bibnamefont {Wang}}, \bibinfo {author} {\bibfnamefont {Y.}~\bibnamefont {Shangguan}}, \bibinfo {author} {\bibfnamefont {Z.}~\bibnamefont {Ma}}, \bibinfo {author} {\bibfnamefont {Z.-Y.}\ \bibnamefont {Dong}}, \bibinfo {author} {\bibfnamefont {R.}~\bibnamefont {Kajimoto}}, \bibinfo {author} {\bibfnamefont {K.}~\bibnamefont {Ikeuchi}}, \bibinfo {author} {\bibfnamefont {S.-L.}\ \bibnamefont {Yu}}, \bibinfo {author} {\bibfnamefont {J.}~\bibnamefont {Sun}}, \bibinfo {author} {\bibfnamefont {J.-X.}\ \bibnamefont {Li}},\ and\ \bibinfo {author} {\bibfnamefont {J.}~\bibnamefont {Wen}},\ }\bibfield  {title} {\bibinfo {title} {Evidence for magnon-phonon coupling in the topological magnet ${\mathrm{cu}}_{3}{\mathrm{teo}}_{6}$},\
  }\href {https://doi.org/10.1103/PhysRevB.101.214419} {\bibfield  {journal} {\bibinfo  {journal} {Phys. Rev. B}\ }\textbf {\bibinfo {volume} {101}},\ \bibinfo {pages} {214419} (\bibinfo {year} {2020})}\BibitemShut {NoStop}%
\bibitem [{\citenamefont {Zhang}\ \emph {et~al.}(2021{\natexlab{a}})\citenamefont {Zhang}, \citenamefont {Xu}, \citenamefont {Carnahan}, \citenamefont {Sretenovic}, \citenamefont {Suri}, \citenamefont {Xiao},\ and\ \citenamefont {Ke}}]{zhang2021anomalous}%
  \BibitemOpen
  \bibfield  {author} {\bibinfo {author} {\bibfnamefont {H.}~\bibnamefont {Zhang}}, \bibinfo {author} {\bibfnamefont {C.}~\bibnamefont {Xu}}, \bibinfo {author} {\bibfnamefont {C.}~\bibnamefont {Carnahan}}, \bibinfo {author} {\bibfnamefont {M.}~\bibnamefont {Sretenovic}}, \bibinfo {author} {\bibfnamefont {N.}~\bibnamefont {Suri}}, \bibinfo {author} {\bibfnamefont {D.}~\bibnamefont {Xiao}},\ and\ \bibinfo {author} {\bibfnamefont {X.}~\bibnamefont {Ke}},\ }\bibfield  {title} {\bibinfo {title} {Anomalous thermal hall effect in an insulating van der waals magnet},\ }\href {https://doi.org/10.1103/PhysRevLett.127.247202} {\bibfield  {journal} {\bibinfo  {journal} {Phys. Rev. Lett.}\ }\textbf {\bibinfo {volume} {127}},\ \bibinfo {pages} {247202} (\bibinfo {year} {2021}{\natexlab{a}})}\BibitemShut {NoStop}%
\bibitem [{\citenamefont {Luo}\ \emph {et~al.}(2023)\citenamefont {Luo}, \citenamefont {Li}, \citenamefont {Ye}, \citenamefont {Xu}, \citenamefont {Yan}, \citenamefont {Zhang}, \citenamefont {Ye}, \citenamefont {Chen}, \citenamefont {Hu}, \citenamefont {Teng} \emph {et~al.}}]{luo2023evidence}%
  \BibitemOpen
  \bibfield  {author} {\bibinfo {author} {\bibfnamefont {J.}~\bibnamefont {Luo}}, \bibinfo {author} {\bibfnamefont {S.}~\bibnamefont {Li}}, \bibinfo {author} {\bibfnamefont {Z.}~\bibnamefont {Ye}}, \bibinfo {author} {\bibfnamefont {R.}~\bibnamefont {Xu}}, \bibinfo {author} {\bibfnamefont {H.}~\bibnamefont {Yan}}, \bibinfo {author} {\bibfnamefont {J.}~\bibnamefont {Zhang}}, \bibinfo {author} {\bibfnamefont {G.}~\bibnamefont {Ye}}, \bibinfo {author} {\bibfnamefont {L.}~\bibnamefont {Chen}}, \bibinfo {author} {\bibfnamefont {D.}~\bibnamefont {Hu}}, \bibinfo {author} {\bibfnamefont {X.}~\bibnamefont {Teng}}, \emph {et~al.},\ }\bibfield  {title} {\bibinfo {title} {Evidence for topological magnon--phonon hybridization in a 2d antiferromagnet down to the monolayer limit},\ }\href {https://doi.org/10.1021/acs.nanolett.3c00351} {\bibfield  {journal} {\bibinfo  {journal} {Nano Letters}\ }\textbf {\bibinfo {volume} {23}},\ \bibinfo {pages} {2023} (\bibinfo {year} {2023})}\BibitemShut {NoStop}%
\bibitem [{\citenamefont {S{\"u}sstrunk}\ and\ \citenamefont {Huber}(2016)}]{susstrunk2016classification}%
  \BibitemOpen
  \bibfield  {author} {\bibinfo {author} {\bibfnamefont {R.}~\bibnamefont {S{\"u}sstrunk}}\ and\ \bibinfo {author} {\bibfnamefont {S.~D.}\ \bibnamefont {Huber}},\ }\bibfield  {title} {\bibinfo {title} {Classification of topological phonons in linear mechanical metamaterials},\ }\href {https://doi.org/10.1073/pnas.1605462113} {\bibfield  {journal} {\bibinfo  {journal} {Proceedings of the National Academy of Sciences}\ }\textbf {\bibinfo {volume} {113}},\ \bibinfo {pages} {E4767} (\bibinfo {year} {2016})}\BibitemShut {NoStop}%
\bibitem [{\citenamefont {Maksimov}\ \emph {et~al.}(2019)\citenamefont {Maksimov}, \citenamefont {Zhu}, \citenamefont {White},\ and\ \citenamefont {Chernyshev}}]{maksimov2019anisotropic}%
  \BibitemOpen
  \bibfield  {author} {\bibinfo {author} {\bibfnamefont {P.~A.}\ \bibnamefont {Maksimov}}, \bibinfo {author} {\bibfnamefont {Z.}~\bibnamefont {Zhu}}, \bibinfo {author} {\bibfnamefont {S.~R.}\ \bibnamefont {White}},\ and\ \bibinfo {author} {\bibfnamefont {A.~L.}\ \bibnamefont {Chernyshev}},\ }\bibfield  {title} {\bibinfo {title} {Anisotropic-exchange magnets on a triangular lattice: Spin waves, accidental degeneracies, and dual spin liquids},\ }\href {https://doi.org/10.1103/PhysRevX.9.021017} {\bibfield  {journal} {\bibinfo  {journal} {Phys. Rev. X}\ }\textbf {\bibinfo {volume} {9}},\ \bibinfo {pages} {021017} (\bibinfo {year} {2019})}\BibitemShut {NoStop}%
\bibitem [{\citenamefont {Holstein}\ and\ \citenamefont {Primakoff}(1940)}]{holstein1940field}%
  \BibitemOpen
  \bibfield  {author} {\bibinfo {author} {\bibfnamefont {T.}~\bibnamefont {Holstein}}\ and\ \bibinfo {author} {\bibfnamefont {H.}~\bibnamefont {Primakoff}},\ }\bibfield  {title} {\bibinfo {title} {Field dependence of the intrinsic domain magnetization of a ferromagnet},\ }\href {https://doi.org/10.1103/PhysRev.58.1098} {\bibfield  {journal} {\bibinfo  {journal} {Phys. Rev.}\ }\textbf {\bibinfo {volume} {58}},\ \bibinfo {pages} {1098} (\bibinfo {year} {1940})}\BibitemShut {NoStop}%
\bibitem [{sm()}]{sm}%
  \BibitemOpen
  \href@noop {} {\bibinfo {title} {See supplementary materials for details on derivations of model hamiltonian, symmetry-adapted magnetoelastic couplings, and the topological classification with the phononic analog.}}\BibitemShut {Stop}%
\bibitem [{\citenamefont {Kittel}(1949)}]{kittel1949physical}%
  \BibitemOpen
  \bibfield  {author} {\bibinfo {author} {\bibfnamefont {C.}~\bibnamefont {Kittel}},\ }\bibfield  {title} {\bibinfo {title} {Physical theory of ferromagnetic domains},\ }\href {https://doi.org/10.1103/RevModPhys.21.541} {\bibfield  {journal} {\bibinfo  {journal} {Rev. Mod. Phys.}\ }\textbf {\bibinfo {volume} {21}},\ \bibinfo {pages} {541} (\bibinfo {year} {1949})}\BibitemShut {NoStop}%
\bibitem [{\citenamefont {Kittel}(1958)}]{kittel1958interaction}%
  \BibitemOpen
  \bibfield  {author} {\bibinfo {author} {\bibfnamefont {C.}~\bibnamefont {Kittel}},\ }\bibfield  {title} {\bibinfo {title} {Interaction of spin waves and ultrasonic waves in ferromagnetic crystals},\ }\href {https://doi.org/10.1103/PhysRev.110.836} {\bibfield  {journal} {\bibinfo  {journal} {Phys. Rev.}\ }\textbf {\bibinfo {volume} {110}},\ \bibinfo {pages} {836} (\bibinfo {year} {1958})}\BibitemShut {NoStop}%
\bibitem [{\citenamefont {Streib}\ \emph {et~al.}(2019)\citenamefont {Streib}, \citenamefont {Vidal-Silva}, \citenamefont {Shen},\ and\ \citenamefont {Bauer}}]{streib2019magnon}%
  \BibitemOpen
  \bibfield  {author} {\bibinfo {author} {\bibfnamefont {S.}~\bibnamefont {Streib}}, \bibinfo {author} {\bibfnamefont {N.}~\bibnamefont {Vidal-Silva}}, \bibinfo {author} {\bibfnamefont {K.}~\bibnamefont {Shen}},\ and\ \bibinfo {author} {\bibfnamefont {G.~E.~W.}\ \bibnamefont {Bauer}},\ }\bibfield  {title} {\bibinfo {title} {Magnon-phonon interactions in magnetic insulators},\ }\href {https://doi.org/10.1103/PhysRevB.99.184442} {\bibfield  {journal} {\bibinfo  {journal} {Phys. Rev. B}\ }\textbf {\bibinfo {volume} {99}},\ \bibinfo {pages} {184442} (\bibinfo {year} {2019})}\BibitemShut {NoStop}%
\bibitem [{\citenamefont {R\"uckriegel}\ \emph {et~al.}(2020)\citenamefont {R\"uckriegel}, \citenamefont {Streib}, \citenamefont {Bauer},\ and\ \citenamefont {Duine}}]{ruckriegel2020angular}%
  \BibitemOpen
  \bibfield  {author} {\bibinfo {author} {\bibfnamefont {A.}~\bibnamefont {R\"uckriegel}}, \bibinfo {author} {\bibfnamefont {S.}~\bibnamefont {Streib}}, \bibinfo {author} {\bibfnamefont {G.~E.~W.}\ \bibnamefont {Bauer}},\ and\ \bibinfo {author} {\bibfnamefont {R.~A.}\ \bibnamefont {Duine}},\ }\bibfield  {title} {\bibinfo {title} {Angular momentum conservation and phonon spin in magnetic insulators},\ }\href {https://doi.org/10.1103/PhysRevB.101.104402} {\bibfield  {journal} {\bibinfo  {journal} {Phys. Rev. B}\ }\textbf {\bibinfo {volume} {101}},\ \bibinfo {pages} {104402} (\bibinfo {year} {2020})}\BibitemShut {NoStop}%
\bibitem [{\citenamefont {Gallego}\ \emph {et~al.}(2019)\citenamefont {Gallego}, \citenamefont {Etxebarria}, \citenamefont {Elcoro}, \citenamefont {Tasci},\ and\ \citenamefont {Perez-Mato}}]{gallego2019automatic}%
  \BibitemOpen
  \bibfield  {author} {\bibinfo {author} {\bibfnamefont {S.~V.}\ \bibnamefont {Gallego}}, \bibinfo {author} {\bibfnamefont {J.}~\bibnamefont {Etxebarria}}, \bibinfo {author} {\bibfnamefont {L.}~\bibnamefont {Elcoro}}, \bibinfo {author} {\bibfnamefont {E.~S.}\ \bibnamefont {Tasci}},\ and\ \bibinfo {author} {\bibfnamefont {J.~M.}\ \bibnamefont {Perez-Mato}},\ }\bibfield  {title} {\bibinfo {title} {Automatic calculation of symmetry-adapted tensors in magnetic and non-magnetic materials: a new tool of the bilbao crystallographic server},\ }\href {https://doi.org/10.1107/S2053273319001748} {\bibfield  {journal} {\bibinfo  {journal} {Acta Crystallographica Section A: Foundations and Advances}\ }\textbf {\bibinfo {volume} {75}},\ \bibinfo {pages} {438} (\bibinfo {year} {2019})}\BibitemShut {NoStop}%
\bibitem [{\citenamefont {Saito}\ \emph {et~al.}(2019)\citenamefont {Saito}, \citenamefont {Misaki}, \citenamefont {Ishizuka},\ and\ \citenamefont {Nagaosa}}]{saito2019berry}%
  \BibitemOpen
  \bibfield  {author} {\bibinfo {author} {\bibfnamefont {T.}~\bibnamefont {Saito}}, \bibinfo {author} {\bibfnamefont {K.}~\bibnamefont {Misaki}}, \bibinfo {author} {\bibfnamefont {H.}~\bibnamefont {Ishizuka}},\ and\ \bibinfo {author} {\bibfnamefont {N.}~\bibnamefont {Nagaosa}},\ }\bibfield  {title} {\bibinfo {title} {Berry phase of phonons and thermal hall effect in nonmagnetic insulators},\ }\href {https://doi.org/10.1103/PhysRevLett.123.255901} {\bibfield  {journal} {\bibinfo  {journal} {Phys. Rev. Lett.}\ }\textbf {\bibinfo {volume} {123}},\ \bibinfo {pages} {255901} (\bibinfo {year} {2019})}\BibitemShut {NoStop}%
\bibitem [{\citenamefont {Berry}(1984)}]{berry1984quantal}%
  \BibitemOpen
  \bibfield  {author} {\bibinfo {author} {\bibfnamefont {M.~V.}\ \bibnamefont {Berry}},\ }\bibfield  {title} {\bibinfo {title} {Quantal phase factors accompanying adiabatic changes},\ }\href {https://doi.org/10.1098/rspa.1984.0023} {\bibfield  {journal} {\bibinfo  {journal} {Proceedings of the Royal Society of London. A. Mathematical and Physical Sciences}\ }\textbf {\bibinfo {volume} {392}},\ \bibinfo {pages} {45} (\bibinfo {year} {1984})}\BibitemShut {NoStop}%
\bibitem [{\citenamefont {Zhang}\ \emph {et~al.}(2010)\citenamefont {Zhang}, \citenamefont {Ren}, \citenamefont {Wang},\ and\ \citenamefont {Li}}]{zhang2010topological}%
  \BibitemOpen
  \bibfield  {author} {\bibinfo {author} {\bibfnamefont {L.}~\bibnamefont {Zhang}}, \bibinfo {author} {\bibfnamefont {J.}~\bibnamefont {Ren}}, \bibinfo {author} {\bibfnamefont {J.-S.}\ \bibnamefont {Wang}},\ and\ \bibinfo {author} {\bibfnamefont {B.}~\bibnamefont {Li}},\ }\bibfield  {title} {\bibinfo {title} {Topological nature of the phonon hall effect},\ }\href {https://doi.org/10.1103/PhysRevLett.105.225901} {\bibfield  {journal} {\bibinfo  {journal} {Phys. Rev. Lett.}\ }\textbf {\bibinfo {volume} {105}},\ \bibinfo {pages} {225901} (\bibinfo {year} {2010})}\BibitemShut {NoStop}%
\bibitem [{\citenamefont {Fukui}\ \emph {et~al.}(2005)\citenamefont {Fukui}, \citenamefont {Hatsugai},\ and\ \citenamefont {Suzuki}}]{fukui2005chern}%
  \BibitemOpen
  \bibfield  {author} {\bibinfo {author} {\bibfnamefont {T.}~\bibnamefont {Fukui}}, \bibinfo {author} {\bibfnamefont {Y.}~\bibnamefont {Hatsugai}},\ and\ \bibinfo {author} {\bibfnamefont {H.}~\bibnamefont {Suzuki}},\ }\bibfield  {title} {\bibinfo {title} {Chern numbers in discretized brillouin zone: efficient method of computing (spin) hall conductances},\ }\href {https://doi.org/10.1143/JPSJ.74.1674} {\bibfield  {journal} {\bibinfo  {journal} {Journal of the Physical Society of Japan}\ }\textbf {\bibinfo {volume} {74}},\ \bibinfo {pages} {1674} (\bibinfo {year} {2005})}\BibitemShut {NoStop}%
\bibitem [{\citenamefont {Strohm}\ \emph {et~al.}(2005)\citenamefont {Strohm}, \citenamefont {Rikken},\ and\ \citenamefont {Wyder}}]{strohm2005phenomenological}%
  \BibitemOpen
  \bibfield  {author} {\bibinfo {author} {\bibfnamefont {C.}~\bibnamefont {Strohm}}, \bibinfo {author} {\bibfnamefont {G.~L. J.~A.}\ \bibnamefont {Rikken}},\ and\ \bibinfo {author} {\bibfnamefont {P.}~\bibnamefont {Wyder}},\ }\bibfield  {title} {\bibinfo {title} {Phenomenological evidence for the phonon hall effect},\ }\href {https://doi.org/10.1103/PhysRevLett.95.155901} {\bibfield  {journal} {\bibinfo  {journal} {Phys. Rev. Lett.}\ }\textbf {\bibinfo {volume} {95}},\ \bibinfo {pages} {155901} (\bibinfo {year} {2005})}\BibitemShut {NoStop}%
\bibitem [{\citenamefont {Matsumoto}\ and\ \citenamefont {Murakami}(2011{\natexlab{a}})}]{matsumoto2011theoretical}%
  \BibitemOpen
  \bibfield  {author} {\bibinfo {author} {\bibfnamefont {R.}~\bibnamefont {Matsumoto}}\ and\ \bibinfo {author} {\bibfnamefont {S.}~\bibnamefont {Murakami}},\ }\bibfield  {title} {\bibinfo {title} {Theoretical prediction of a rotating magnon wave packet in ferromagnets},\ }\href {https://doi.org/10.1103/PhysRevLett.106.197202} {\bibfield  {journal} {\bibinfo  {journal} {Phys. Rev. Lett.}\ }\textbf {\bibinfo {volume} {106}},\ \bibinfo {pages} {197202} (\bibinfo {year} {2011}{\natexlab{a}})}\BibitemShut {NoStop}%
\bibitem [{\citenamefont {Sugii}\ \emph {et~al.}(2017)\citenamefont {Sugii}, \citenamefont {Shimozawa}, \citenamefont {Watanabe}, \citenamefont {Suzuki}, \citenamefont {Halim}, \citenamefont {Kimata}, \citenamefont {Matsumoto}, \citenamefont {Nakatsuji},\ and\ \citenamefont {Yamashita}}]{sugii2017thermal}%
  \BibitemOpen
  \bibfield  {author} {\bibinfo {author} {\bibfnamefont {K.}~\bibnamefont {Sugii}}, \bibinfo {author} {\bibfnamefont {M.}~\bibnamefont {Shimozawa}}, \bibinfo {author} {\bibfnamefont {D.}~\bibnamefont {Watanabe}}, \bibinfo {author} {\bibfnamefont {Y.}~\bibnamefont {Suzuki}}, \bibinfo {author} {\bibfnamefont {M.}~\bibnamefont {Halim}}, \bibinfo {author} {\bibfnamefont {M.}~\bibnamefont {Kimata}}, \bibinfo {author} {\bibfnamefont {Y.}~\bibnamefont {Matsumoto}}, \bibinfo {author} {\bibfnamefont {S.}~\bibnamefont {Nakatsuji}},\ and\ \bibinfo {author} {\bibfnamefont {M.}~\bibnamefont {Yamashita}},\ }\bibfield  {title} {\bibinfo {title} {Thermal hall effect in a phonon-glass ${\mathrm{ba}}_{3}{\mathrm{cusb}}_{2}{\mathrm{o}}_{9}$},\ }\href {https://doi.org/10.1103/PhysRevLett.118.145902} {\bibfield  {journal} {\bibinfo  {journal} {Phys. Rev. Lett.}\ }\textbf {\bibinfo {volume} {118}},\ \bibinfo {pages} {145902} (\bibinfo {year} {2017})}\BibitemShut {NoStop}%
\bibitem [{\citenamefont {Grissonnanche}\ \emph {et~al.}(2020)\citenamefont {Grissonnanche}, \citenamefont {Th{\'e}riault}, \citenamefont {Gourgout}, \citenamefont {Boulanger}, \citenamefont {Lefran{\c{c}}ois}, \citenamefont {Ataei}, \citenamefont {Lalibert{\'e}}, \citenamefont {Dion}, \citenamefont {Zhou}, \citenamefont {Pyon} \emph {et~al.}}]{grissonnanche2020chiral}%
  \BibitemOpen
  \bibfield  {author} {\bibinfo {author} {\bibfnamefont {G.}~\bibnamefont {Grissonnanche}}, \bibinfo {author} {\bibfnamefont {S.}~\bibnamefont {Th{\'e}riault}}, \bibinfo {author} {\bibfnamefont {A.}~\bibnamefont {Gourgout}}, \bibinfo {author} {\bibfnamefont {M.-E.}\ \bibnamefont {Boulanger}}, \bibinfo {author} {\bibfnamefont {E.}~\bibnamefont {Lefran{\c{c}}ois}}, \bibinfo {author} {\bibfnamefont {A.}~\bibnamefont {Ataei}}, \bibinfo {author} {\bibfnamefont {F.}~\bibnamefont {Lalibert{\'e}}}, \bibinfo {author} {\bibfnamefont {M.}~\bibnamefont {Dion}}, \bibinfo {author} {\bibfnamefont {J.-S.}\ \bibnamefont {Zhou}}, \bibinfo {author} {\bibfnamefont {S.}~\bibnamefont {Pyon}}, \emph {et~al.},\ }\bibfield  {title} {\bibinfo {title} {Chiral phonons in the pseudogap phase of cuprates},\ }\href {https://doi.org/10.1038/s41567-020-0965-y} {\bibfield  {journal} {\bibinfo  {journal} {Nature Physics}\ }\textbf {\bibinfo {volume} {16}},\ \bibinfo {pages} {1108} (\bibinfo {year} {2020})}\BibitemShut {NoStop}%
\bibitem [{\citenamefont {Zhang}\ \emph {et~al.}(2024)\citenamefont {Zhang}, \citenamefont {Gao},\ and\ \citenamefont {Chen}}]{zhang2024thermal}%
  \BibitemOpen
  \bibfield  {author} {\bibinfo {author} {\bibfnamefont {X.-T.}\ \bibnamefont {Zhang}}, \bibinfo {author} {\bibfnamefont {Y.~H.}\ \bibnamefont {Gao}},\ and\ \bibinfo {author} {\bibfnamefont {G.}~\bibnamefont {Chen}},\ }\bibfield  {title} {\bibinfo {title} {Thermal hall effects in quantum magnets},\ }\href {https://doi.org/10.1016/j.physrep.2024.03.004} {\bibfield  {journal} {\bibinfo  {journal} {Physics Reports}\ }\textbf {\bibinfo {volume} {1070}},\ \bibinfo {pages} {1} (\bibinfo {year} {2024})}\BibitemShut {NoStop}%
\bibitem [{\citenamefont {Matsumoto}\ and\ \citenamefont {Murakami}(2011{\natexlab{b}})}]{matsumoto2011rotational}%
  \BibitemOpen
  \bibfield  {author} {\bibinfo {author} {\bibfnamefont {R.}~\bibnamefont {Matsumoto}}\ and\ \bibinfo {author} {\bibfnamefont {S.}~\bibnamefont {Murakami}},\ }\bibfield  {title} {\bibinfo {title} {Rotational motion of magnons and the thermal hall effect},\ }\href {https://doi.org/10.1103/PhysRevB.84.184406} {\bibfield  {journal} {\bibinfo  {journal} {Phys. Rev. B}\ }\textbf {\bibinfo {volume} {84}},\ \bibinfo {pages} {184406} (\bibinfo {year} {2011}{\natexlab{b}})}\BibitemShut {NoStop}%
\bibitem [{\citenamefont {Li}\ \emph {et~al.}(2020)\citenamefont {Li}, \citenamefont {Fauqu\'e}, \citenamefont {Zhu},\ and\ \citenamefont {Behnia}}]{li2020phonon}%
  \BibitemOpen
  \bibfield  {author} {\bibinfo {author} {\bibfnamefont {X.}~\bibnamefont {Li}}, \bibinfo {author} {\bibfnamefont {B.}~\bibnamefont {Fauqu\'e}}, \bibinfo {author} {\bibfnamefont {Z.}~\bibnamefont {Zhu}},\ and\ \bibinfo {author} {\bibfnamefont {K.}~\bibnamefont {Behnia}},\ }\bibfield  {title} {\bibinfo {title} {Phonon thermal hall effect in strontium titanate},\ }\href {https://doi.org/10.1103/PhysRevLett.124.105901} {\bibfield  {journal} {\bibinfo  {journal} {Phys. Rev. Lett.}\ }\textbf {\bibinfo {volume} {124}},\ \bibinfo {pages} {105901} (\bibinfo {year} {2020})}\BibitemShut {NoStop}%
\bibitem [{\citenamefont {Chen}\ \emph {et~al.}(2022)\citenamefont {Chen}, \citenamefont {Boulanger}, \citenamefont {Wang}, \citenamefont {Tafti},\ and\ \citenamefont {Taillefer}}]{chen2022large}%
  \BibitemOpen
  \bibfield  {author} {\bibinfo {author} {\bibfnamefont {L.}~\bibnamefont {Chen}}, \bibinfo {author} {\bibfnamefont {M.-E.}\ \bibnamefont {Boulanger}}, \bibinfo {author} {\bibfnamefont {Z.-C.}\ \bibnamefont {Wang}}, \bibinfo {author} {\bibfnamefont {F.}~\bibnamefont {Tafti}},\ and\ \bibinfo {author} {\bibfnamefont {L.}~\bibnamefont {Taillefer}},\ }\bibfield  {title} {\bibinfo {title} {Large phonon thermal hall conductivity in the antiferromagnetic insulator cu3teo6},\ }\href {https://doi.org/10.1073/pnas.2208016119} {\bibfield  {journal} {\bibinfo  {journal} {Proceedings of the National Academy of Sciences}\ }\textbf {\bibinfo {volume} {119}},\ \bibinfo {pages} {e2208016119} (\bibinfo {year} {2022})}\BibitemShut {NoStop}%
\bibitem [{\citenamefont {Li}\ \emph {et~al.}(2023)\citenamefont {Li}, \citenamefont {Machida}, \citenamefont {Subedi}, \citenamefont {Zhu}, \citenamefont {Li},\ and\ \citenamefont {Behnia}}]{li2023phonon}%
  \BibitemOpen
  \bibfield  {author} {\bibinfo {author} {\bibfnamefont {X.}~\bibnamefont {Li}}, \bibinfo {author} {\bibfnamefont {Y.}~\bibnamefont {Machida}}, \bibinfo {author} {\bibfnamefont {A.}~\bibnamefont {Subedi}}, \bibinfo {author} {\bibfnamefont {Z.}~\bibnamefont {Zhu}}, \bibinfo {author} {\bibfnamefont {L.}~\bibnamefont {Li}},\ and\ \bibinfo {author} {\bibfnamefont {K.}~\bibnamefont {Behnia}},\ }\bibfield  {title} {\bibinfo {title} {The phonon thermal hall angle in black phosphorus},\ }\href {https://doi.org/10.1038/s41467-023-36750-3} {\bibfield  {journal} {\bibinfo  {journal} {Nature Communications}\ }\textbf {\bibinfo {volume} {14}},\ \bibinfo {pages} {1027} (\bibinfo {year} {2023})}\BibitemShut {NoStop}%
\bibitem [{\citenamefont {Onose}\ \emph {et~al.}(2010)\citenamefont {Onose}, \citenamefont {Ideue}, \citenamefont {Katsura}, \citenamefont {Shiomi}, \citenamefont {Nagaosa},\ and\ \citenamefont {Tokura}}]{onose2010observation}%
  \BibitemOpen
  \bibfield  {author} {\bibinfo {author} {\bibfnamefont {Y.}~\bibnamefont {Onose}}, \bibinfo {author} {\bibfnamefont {T.}~\bibnamefont {Ideue}}, \bibinfo {author} {\bibfnamefont {H.}~\bibnamefont {Katsura}}, \bibinfo {author} {\bibfnamefont {Y.}~\bibnamefont {Shiomi}}, \bibinfo {author} {\bibfnamefont {N.}~\bibnamefont {Nagaosa}},\ and\ \bibinfo {author} {\bibfnamefont {Y.}~\bibnamefont {Tokura}},\ }\bibfield  {title} {\bibinfo {title} {Observation of the magnon hall effect},\ }\href {https://doi.org/10.1126/science.1188260} {\bibfield  {journal} {\bibinfo  {journal} {Science}\ }\textbf {\bibinfo {volume} {329}},\ \bibinfo {pages} {297} (\bibinfo {year} {2010})}\BibitemShut {NoStop}%
\bibitem [{\citenamefont {Hirschberger}\ \emph {et~al.}(2015)\citenamefont {Hirschberger}, \citenamefont {Krizan}, \citenamefont {Cava},\ and\ \citenamefont {Ong}}]{hirschberger2015large}%
  \BibitemOpen
  \bibfield  {author} {\bibinfo {author} {\bibfnamefont {M.}~\bibnamefont {Hirschberger}}, \bibinfo {author} {\bibfnamefont {J.~W.}\ \bibnamefont {Krizan}}, \bibinfo {author} {\bibfnamefont {R.}~\bibnamefont {Cava}},\ and\ \bibinfo {author} {\bibfnamefont {N.}~\bibnamefont {Ong}},\ }\bibfield  {title} {\bibinfo {title} {Large thermal hall conductivity of neutral spin excitations in a frustrated quantum magnet},\ }\href {https://doi.org/10.1126/science.1257340} {\bibfield  {journal} {\bibinfo  {journal} {Science}\ }\textbf {\bibinfo {volume} {348}},\ \bibinfo {pages} {106} (\bibinfo {year} {2015})}\BibitemShut {NoStop}%
\bibitem [{\citenamefont {Banerjee}\ \emph {et~al.}(2018)\citenamefont {Banerjee}, \citenamefont {Heiblum}, \citenamefont {Umansky}, \citenamefont {Feldman}, \citenamefont {Oreg},\ and\ \citenamefont {Stern}}]{banerjee2018observation}%
  \BibitemOpen
  \bibfield  {author} {\bibinfo {author} {\bibfnamefont {M.}~\bibnamefont {Banerjee}}, \bibinfo {author} {\bibfnamefont {M.}~\bibnamefont {Heiblum}}, \bibinfo {author} {\bibfnamefont {V.}~\bibnamefont {Umansky}}, \bibinfo {author} {\bibfnamefont {D.~E.}\ \bibnamefont {Feldman}}, \bibinfo {author} {\bibfnamefont {Y.}~\bibnamefont {Oreg}},\ and\ \bibinfo {author} {\bibfnamefont {A.}~\bibnamefont {Stern}},\ }\bibfield  {title} {\bibinfo {title} {Observation of half-integer thermal hall conductance},\ }\href {https://doi.org/10.1038/s41586-018-0184-1} {\bibfield  {journal} {\bibinfo  {journal} {Nature}\ }\textbf {\bibinfo {volume} {559}},\ \bibinfo {pages} {205} (\bibinfo {year} {2018})}\BibitemShut {NoStop}%
\bibitem [{\citenamefont {Park}\ and\ \citenamefont {Yang}(2020)}]{park2020phonon}%
  \BibitemOpen
  \bibfield  {author} {\bibinfo {author} {\bibfnamefont {S.}~\bibnamefont {Park}}\ and\ \bibinfo {author} {\bibfnamefont {B.-J.}\ \bibnamefont {Yang}},\ }\bibfield  {title} {\bibinfo {title} {Phonon angular momentum hall effect},\ }\href {https://doi.org/10.1021/acs.nanolett.0c03220} {\bibfield  {journal} {\bibinfo  {journal} {Nano Letters}\ }\textbf {\bibinfo {volume} {20}},\ \bibinfo {pages} {7694} (\bibinfo {year} {2020})}\BibitemShut {NoStop}%
\bibitem [{\citenamefont {Pocs}\ \emph {et~al.}(2024)\citenamefont {Pocs}, \citenamefont {Xing}, \citenamefont {Choi}, \citenamefont {Sefat}, \citenamefont {Hermele},\ and\ \citenamefont {Lee}}]{pocs2024generic}%
  \BibitemOpen
  \bibfield  {author} {\bibinfo {author} {\bibfnamefont {C.~A.}\ \bibnamefont {Pocs}}, \bibinfo {author} {\bibfnamefont {J.}~\bibnamefont {Xing}}, \bibinfo {author} {\bibfnamefont {E.~S.}\ \bibnamefont {Choi}}, \bibinfo {author} {\bibfnamefont {A.~S.}\ \bibnamefont {Sefat}}, \bibinfo {author} {\bibfnamefont {M.}~\bibnamefont {Hermele}},\ and\ \bibinfo {author} {\bibfnamefont {M.}~\bibnamefont {Lee}},\ }\bibfield  {title} {\bibinfo {title} {Generic magnetic field dependence of thermal conductivity in effective spin-1/2 magnetic insulators via hybridization of acoustic phonons and spin-flip excitations},\ }\href {https://arxiv.org/abs/2401.01407} {\bibfield  {journal} {\bibinfo  {journal} {arXiv preprint arXiv:2401.01407}\ } (\bibinfo {year} {2024})}\BibitemShut {NoStop}%
\bibitem [{\citenamefont {Del~Maestro}\ and\ \citenamefont {Gingras}(2004)}]{del2004quantum}%
  \BibitemOpen
  \bibfield  {author} {\bibinfo {author} {\bibfnamefont {A.~G.}\ \bibnamefont {Del~Maestro}}\ and\ \bibinfo {author} {\bibfnamefont {M.~J.}\ \bibnamefont {Gingras}},\ }\bibfield  {title} {\bibinfo {title} {Quantum spin fluctuations in the dipolar heisenberg-like rare earth pyrochlores},\ }\href {https://doi.org/10.1088/0953-8984/16/20/005} {\bibfield  {journal} {\bibinfo  {journal} {Journal of Physics: Condensed Matter}\ }\textbf {\bibinfo {volume} {16}},\ \bibinfo {pages} {3339} (\bibinfo {year} {2004})}\BibitemShut {NoStop}%
\bibitem [{\citenamefont {Li}\ \emph {et~al.}(2016)\citenamefont {Li}, \citenamefont {Wang},\ and\ \citenamefont {Chen}}]{PhysRevB.94.035107}%
  \BibitemOpen
  \bibfield  {author} {\bibinfo {author} {\bibfnamefont {Y.-D.}\ \bibnamefont {Li}}, \bibinfo {author} {\bibfnamefont {X.}~\bibnamefont {Wang}},\ and\ \bibinfo {author} {\bibfnamefont {G.}~\bibnamefont {Chen}},\ }\bibfield  {title} {\bibinfo {title} {Anisotropic spin model of strong spin-orbit-coupled triangular antiferromagnets},\ }\href {https://doi.org/10.1103/PhysRevB.94.035107} {\bibfield  {journal} {\bibinfo  {journal} {Phys. Rev. B}\ }\textbf {\bibinfo {volume} {94}},\ \bibinfo {pages} {035107} (\bibinfo {year} {2016})}\BibitemShut {NoStop}%
\bibitem [{\citenamefont {Liu}\ \emph {et~al.}(2018)\citenamefont {Liu}, \citenamefont {Zhang}, \citenamefont {Ji}, \citenamefont {Liu}, \citenamefont {Li}, \citenamefont {Wang}, \citenamefont {Lei}, \citenamefont {Chen},\ and\ \citenamefont {Zhang}}]{Liu_2018}%
  \BibitemOpen
  \bibfield  {author} {\bibinfo {author} {\bibfnamefont {W.}~\bibnamefont {Liu}}, \bibinfo {author} {\bibfnamefont {Z.}~\bibnamefont {Zhang}}, \bibinfo {author} {\bibfnamefont {J.}~\bibnamefont {Ji}}, \bibinfo {author} {\bibfnamefont {Y.}~\bibnamefont {Liu}}, \bibinfo {author} {\bibfnamefont {J.}~\bibnamefont {Li}}, \bibinfo {author} {\bibfnamefont {X.}~\bibnamefont {Wang}}, \bibinfo {author} {\bibfnamefont {H.}~\bibnamefont {Lei}}, \bibinfo {author} {\bibfnamefont {G.}~\bibnamefont {Chen}},\ and\ \bibinfo {author} {\bibfnamefont {Q.}~\bibnamefont {Zhang}},\ }\bibfield  {title} {\bibinfo {title} {Rare-earth chalcogenides: A large family of triangular lattice spin liquid candidates},\ }\href {https://doi.org/10.1088/0256-307x/35/11/117501} {\bibfield  {journal} {\bibinfo  {journal} {Chinese Physics Letters}\ }\textbf {\bibinfo {volume} {35}},\ \bibinfo {pages} {117501} (\bibinfo {year} {2018})}\BibitemShut {NoStop}%
\bibitem [{\citenamefont {Bai}\ \emph {et~al.}(2021)\citenamefont {Bai}, \citenamefont {Zhang}, \citenamefont {Dun}, \citenamefont {Zhang}, \citenamefont {Huang}, \citenamefont {Zhou}, \citenamefont {Stone}, \citenamefont {Kolesnikov}, \citenamefont {Ye}, \citenamefont {Batista},\ and\ \citenamefont {Mourigal}}]{Bai_2021}%
  \BibitemOpen
  \bibfield  {author} {\bibinfo {author} {\bibfnamefont {X.}~\bibnamefont {Bai}}, \bibinfo {author} {\bibfnamefont {S.-S.}\ \bibnamefont {Zhang}}, \bibinfo {author} {\bibfnamefont {Z.}~\bibnamefont {Dun}}, \bibinfo {author} {\bibfnamefont {H.}~\bibnamefont {Zhang}}, \bibinfo {author} {\bibfnamefont {Q.}~\bibnamefont {Huang}}, \bibinfo {author} {\bibfnamefont {H.}~\bibnamefont {Zhou}}, \bibinfo {author} {\bibfnamefont {M.~B.}\ \bibnamefont {Stone}}, \bibinfo {author} {\bibfnamefont {A.~I.}\ \bibnamefont {Kolesnikov}}, \bibinfo {author} {\bibfnamefont {F.}~\bibnamefont {Ye}}, \bibinfo {author} {\bibfnamefont {C.~D.}\ \bibnamefont {Batista}},\ and\ \bibinfo {author} {\bibfnamefont {M.}~\bibnamefont {Mourigal}},\ }\bibfield  {title} {\bibinfo {title} {Hybridized quadrupolar excitations in the spin-anisotropic frustrated magnet {FeI}2},\ }\href {https://doi.org/10.1038/s41567-020-01110-1} {\bibfield  {journal} {\bibinfo  {journal} {Nature Physics}\ }\textbf {\bibinfo {volume} {17}},\ \bibinfo {pages} {467}
  (\bibinfo {year} {2021})}\BibitemShut {NoStop}%
\bibitem [{\citenamefont {Legros}\ \emph {et~al.}(2021)\citenamefont {Legros}, \citenamefont {Zhang}, \citenamefont {Bai}, \citenamefont {Zhang}, \citenamefont {Dun}, \citenamefont {Phelan}, \citenamefont {Batista}, \citenamefont {Mourigal},\ and\ \citenamefont {Armitage}}]{PhysRevLett.127.267201}%
  \BibitemOpen
  \bibfield  {author} {\bibinfo {author} {\bibfnamefont {A.}~\bibnamefont {Legros}}, \bibinfo {author} {\bibfnamefont {S.-S.}\ \bibnamefont {Zhang}}, \bibinfo {author} {\bibfnamefont {X.}~\bibnamefont {Bai}}, \bibinfo {author} {\bibfnamefont {H.}~\bibnamefont {Zhang}}, \bibinfo {author} {\bibfnamefont {Z.}~\bibnamefont {Dun}}, \bibinfo {author} {\bibfnamefont {W.~A.}\ \bibnamefont {Phelan}}, \bibinfo {author} {\bibfnamefont {C.~D.}\ \bibnamefont {Batista}}, \bibinfo {author} {\bibfnamefont {M.}~\bibnamefont {Mourigal}},\ and\ \bibinfo {author} {\bibfnamefont {N.~P.}\ \bibnamefont {Armitage}},\ }\bibfield  {title} {\bibinfo {title} {{Observation of 4- and 6-Magnon Bound States in the Spin-Anisotropic Frustrated Antiferromagnet ${\mathrm{FeI}}_{2}$}},\ }\href {https://doi.org/10.1103/PhysRevLett.127.267201} {\bibfield  {journal} {\bibinfo  {journal} {Phys. Rev. Lett.}\ }\textbf {\bibinfo {volume} {127}},\ \bibinfo {pages} {267201} (\bibinfo {year} {2021})}\BibitemShut {NoStop}%
\bibitem [{\citenamefont {Chen}(2023)}]{chen2023quadrupole}%
  \BibitemOpen
  \bibfield  {author} {\bibinfo {author} {\bibfnamefont {G.}~\bibnamefont {Chen}},\ }\bibfield  {title} {\bibinfo {title} {Quadrupole moments and their interactions in the triangular lattice antiferromagnet ${\mathrm{fei}}_{2}$},\ }\href {https://doi.org/10.1103/PhysRevResearch.5.L032042} {\bibfield  {journal} {\bibinfo  {journal} {Phys. Rev. Res.}\ }\textbf {\bibinfo {volume} {5}},\ \bibinfo {pages} {L032042} (\bibinfo {year} {2023})}\BibitemShut {NoStop}%
\bibitem [{\citenamefont {Kim}\ \emph {et~al.}(2023)\citenamefont {Kim}, \citenamefont {Kim}, \citenamefont {Park}, \citenamefont {Kim}, \citenamefont {Jeong}, \citenamefont {Ohira-Kawamura}, \citenamefont {Murai}, \citenamefont {Nakajima}, \citenamefont {Chernyshev}, \citenamefont {Mourigal} \emph {et~al.}}]{kim2023bond}%
  \BibitemOpen
  \bibfield  {author} {\bibinfo {author} {\bibfnamefont {C.}~\bibnamefont {Kim}}, \bibinfo {author} {\bibfnamefont {S.}~\bibnamefont {Kim}}, \bibinfo {author} {\bibfnamefont {P.}~\bibnamefont {Park}}, \bibinfo {author} {\bibfnamefont {T.}~\bibnamefont {Kim}}, \bibinfo {author} {\bibfnamefont {J.}~\bibnamefont {Jeong}}, \bibinfo {author} {\bibfnamefont {S.}~\bibnamefont {Ohira-Kawamura}}, \bibinfo {author} {\bibfnamefont {N.}~\bibnamefont {Murai}}, \bibinfo {author} {\bibfnamefont {K.}~\bibnamefont {Nakajima}}, \bibinfo {author} {\bibfnamefont {A.}~\bibnamefont {Chernyshev}}, \bibinfo {author} {\bibfnamefont {M.}~\bibnamefont {Mourigal}}, \emph {et~al.},\ }\bibfield  {title} {\bibinfo {title} {Bond-dependent anisotropy and magnon decay in cobalt-based kitaev triangular antiferromagnet},\ }\href {https://doi.org/10.1038/s41567-023-02180-7} {\bibfield  {journal} {\bibinfo  {journal} {Nature Physics}\ }\textbf {\bibinfo {volume} {19}},\ \bibinfo {pages} {1624} (\bibinfo {year} {2023})}\BibitemShut {NoStop}%
\bibitem [{\citenamefont {Trebst}\ and\ \citenamefont {Hickey}(2022)}]{TREBST20221}%
  \BibitemOpen
  \bibfield  {author} {\bibinfo {author} {\bibfnamefont {S.}~\bibnamefont {Trebst}}\ and\ \bibinfo {author} {\bibfnamefont {C.}~\bibnamefont {Hickey}},\ }\bibfield  {title} {\bibinfo {title} {Kitaev materials},\ }\href {https://doi.org/https://doi.org/10.1016/j.physrep.2021.11.003} {\bibfield  {journal} {\bibinfo  {journal} {Physics Reports}\ }\textbf {\bibinfo {volume} {950}},\ \bibinfo {pages} {1} (\bibinfo {year} {2022})},\ \bibinfo {note} {kitaev materials}\BibitemShut {NoStop}%
\bibitem [{\citenamefont {Rau}\ \emph {et~al.}(2014)\citenamefont {Rau}, \citenamefont {Lee},\ and\ \citenamefont {Kee}}]{rau2014generic}%
  \BibitemOpen
  \bibfield  {author} {\bibinfo {author} {\bibfnamefont {J.~G.}\ \bibnamefont {Rau}}, \bibinfo {author} {\bibfnamefont {E.~K.-H.}\ \bibnamefont {Lee}},\ and\ \bibinfo {author} {\bibfnamefont {H.-Y.}\ \bibnamefont {Kee}},\ }\bibfield  {title} {\bibinfo {title} {Generic spin model for the honeycomb iridates beyond the kitaev limit},\ }\href {https://doi.org/10.1103/PhysRevLett.112.077204} {\bibfield  {journal} {\bibinfo  {journal} {Phys. Rev. Lett.}\ }\textbf {\bibinfo {volume} {112}},\ \bibinfo {pages} {077204} (\bibinfo {year} {2014})}\BibitemShut {NoStop}%
\bibitem [{\citenamefont {Kocsis}\ \emph {et~al.}(2022)\citenamefont {Kocsis}, \citenamefont {Kaib}, \citenamefont {Riedl}, \citenamefont {Gass}, \citenamefont {Lampen-Kelley}, \citenamefont {Mandrus}, \citenamefont {Nagler}, \citenamefont {P\'erez}, \citenamefont {Nielsch}, \citenamefont {B\"uchner}, \citenamefont {Wolter},\ and\ \citenamefont {Valent\'{\i}}}]{kocsis2022magnetoelastic}%
  \BibitemOpen
  \bibfield  {author} {\bibinfo {author} {\bibfnamefont {V.}~\bibnamefont {Kocsis}}, \bibinfo {author} {\bibfnamefont {D.~A.~S.}\ \bibnamefont {Kaib}}, \bibinfo {author} {\bibfnamefont {K.}~\bibnamefont {Riedl}}, \bibinfo {author} {\bibfnamefont {S.}~\bibnamefont {Gass}}, \bibinfo {author} {\bibfnamefont {P.}~\bibnamefont {Lampen-Kelley}}, \bibinfo {author} {\bibfnamefont {D.~G.}\ \bibnamefont {Mandrus}}, \bibinfo {author} {\bibfnamefont {S.~E.}\ \bibnamefont {Nagler}}, \bibinfo {author} {\bibfnamefont {N.}~\bibnamefont {P\'erez}}, \bibinfo {author} {\bibfnamefont {K.}~\bibnamefont {Nielsch}}, \bibinfo {author} {\bibfnamefont {B.}~\bibnamefont {B\"uchner}}, \bibinfo {author} {\bibfnamefont {A.~U.~B.}\ \bibnamefont {Wolter}},\ and\ \bibinfo {author} {\bibfnamefont {R.}~\bibnamefont {Valent\'{\i}}},\ }\bibfield  {title} {\bibinfo {title} {Magnetoelastic coupling anisotropy in the kitaev material $\ensuremath{\alpha}\text{\ensuremath{-}}\mathrm{Ru}{\mathrm{cl}}_{3}$},\ }\href
  {https://doi.org/10.1103/PhysRevB.105.094410} {\bibfield  {journal} {\bibinfo  {journal} {Phys. Rev. B}\ }\textbf {\bibinfo {volume} {105}},\ \bibinfo {pages} {094410} (\bibinfo {year} {2022})}\BibitemShut {NoStop}%
\bibitem [{\citenamefont {Chaloupka}\ \emph {et~al.}(2013)\citenamefont {Chaloupka}, \citenamefont {Jackeli},\ and\ \citenamefont {Khaliullin}}]{chaloupka2013zigzag}%
  \BibitemOpen
  \bibfield  {author} {\bibinfo {author} {\bibfnamefont {J.~c.~v.}\ \bibnamefont {Chaloupka}}, \bibinfo {author} {\bibfnamefont {G.}~\bibnamefont {Jackeli}},\ and\ \bibinfo {author} {\bibfnamefont {G.}~\bibnamefont {Khaliullin}},\ }\bibfield  {title} {\bibinfo {title} {Zigzag magnetic order in the iridium oxide ${\mathrm{na}}_{2}{\mathrm{iro}}_{3}$},\ }\href {https://doi.org/10.1103/PhysRevLett.110.097204} {\bibfield  {journal} {\bibinfo  {journal} {Phys. Rev. Lett.}\ }\textbf {\bibinfo {volume} {110}},\ \bibinfo {pages} {097204} (\bibinfo {year} {2013})}\BibitemShut {NoStop}%
\bibitem [{\citenamefont {Kitaev}(2006)}]{kitaev2006anyons}%
  \BibitemOpen
  \bibfield  {author} {\bibinfo {author} {\bibfnamefont {A.}~\bibnamefont {Kitaev}},\ }\bibfield  {title} {\bibinfo {title} {Anyons in an exactly solved model and beyond},\ }\href {https://doi.org/10.1016/j.aop.2005.10.005} {\bibfield  {journal} {\bibinfo  {journal} {Annals of Physics}\ }\textbf {\bibinfo {volume} {321}},\ \bibinfo {pages} {2} (\bibinfo {year} {2006})}\BibitemShut {NoStop}%
\bibitem [{\citenamefont {Takagi}\ \emph {et~al.}(2019)\citenamefont {Takagi}, \citenamefont {Takayama}, \citenamefont {Jackeli}, \citenamefont {Khaliullin},\ and\ \citenamefont {Nagler}}]{takagi2019concept}%
  \BibitemOpen
  \bibfield  {author} {\bibinfo {author} {\bibfnamefont {H.}~\bibnamefont {Takagi}}, \bibinfo {author} {\bibfnamefont {T.}~\bibnamefont {Takayama}}, \bibinfo {author} {\bibfnamefont {G.}~\bibnamefont {Jackeli}}, \bibinfo {author} {\bibfnamefont {G.}~\bibnamefont {Khaliullin}},\ and\ \bibinfo {author} {\bibfnamefont {S.~E.}\ \bibnamefont {Nagler}},\ }\bibfield  {title} {\bibinfo {title} {Concept and realization of kitaev quantum spin liquids},\ }\href {https://doi.org/10.1038/s42254-019-0038-2} {\bibfield  {journal} {\bibinfo  {journal} {Nature Reviews Physics}\ }\textbf {\bibinfo {volume} {1}},\ \bibinfo {pages} {264} (\bibinfo {year} {2019})}\BibitemShut {NoStop}%
\bibitem [{\citenamefont {Zhang}\ \emph {et~al.}(2021{\natexlab{b}})\citenamefont {Zhang}, \citenamefont {Teng}, \citenamefont {Samajdar}, \citenamefont {Sachdev},\ and\ \citenamefont {Scheurer}}]{zhang2021phonon}%
  \BibitemOpen
  \bibfield  {author} {\bibinfo {author} {\bibfnamefont {Y.}~\bibnamefont {Zhang}}, \bibinfo {author} {\bibfnamefont {Y.}~\bibnamefont {Teng}}, \bibinfo {author} {\bibfnamefont {R.}~\bibnamefont {Samajdar}}, \bibinfo {author} {\bibfnamefont {S.}~\bibnamefont {Sachdev}},\ and\ \bibinfo {author} {\bibfnamefont {M.~S.}\ \bibnamefont {Scheurer}},\ }\bibfield  {title} {\bibinfo {title} {Phonon hall viscosity from phonon-spinon interactions},\ }\href {https://doi.org/10.1103/PhysRevB.104.035103} {\bibfield  {journal} {\bibinfo  {journal} {Phys. Rev. B}\ }\textbf {\bibinfo {volume} {104}},\ \bibinfo {pages} {035103} (\bibinfo {year} {2021}{\natexlab{b}})}\BibitemShut {NoStop}%
\bibitem [{\citenamefont {Cui}\ \emph {et~al.}(2023)\citenamefont {Cui}, \citenamefont {Bostr{\"o}m}, \citenamefont {Ozerov}, \citenamefont {Wu}, \citenamefont {Jiang}, \citenamefont {Chu}, \citenamefont {Li}, \citenamefont {Liu}, \citenamefont {Xu}, \citenamefont {Rubio} \emph {et~al.}}]{cui2023chirality}%
  \BibitemOpen
  \bibfield  {author} {\bibinfo {author} {\bibfnamefont {J.}~\bibnamefont {Cui}}, \bibinfo {author} {\bibfnamefont {E.~V.}\ \bibnamefont {Bostr{\"o}m}}, \bibinfo {author} {\bibfnamefont {M.}~\bibnamefont {Ozerov}}, \bibinfo {author} {\bibfnamefont {F.}~\bibnamefont {Wu}}, \bibinfo {author} {\bibfnamefont {Q.}~\bibnamefont {Jiang}}, \bibinfo {author} {\bibfnamefont {J.-H.}\ \bibnamefont {Chu}}, \bibinfo {author} {\bibfnamefont {C.}~\bibnamefont {Li}}, \bibinfo {author} {\bibfnamefont {F.}~\bibnamefont {Liu}}, \bibinfo {author} {\bibfnamefont {X.}~\bibnamefont {Xu}}, \bibinfo {author} {\bibfnamefont {A.}~\bibnamefont {Rubio}}, \emph {et~al.},\ }\bibfield  {title} {\bibinfo {title} {Chirality selective magnon-phonon hybridization and magnon-induced chiral phonons in a layered zigzag antiferromagnet},\ }\href {https://doi.org/10.1038/s41467-023-39123-y} {\bibfield  {journal} {\bibinfo  {journal} {Nature Communications}\ }\textbf {\bibinfo {volume} {14}},\ \bibinfo {pages} {3396} (\bibinfo {year}
  {2023})}\BibitemShut {NoStop}%
\end{thebibliography}%

\widetext
\clearpage
\begin{center}
\textbf{\large Supplementary Materials for ``Chiral phonons induced from spin dynamics via magnetoelastic anisotropy''}
\end{center}
\begin{center}
    Bowen Ma$^{1}$, Z. D. Wang$^{1,2}$, and Gang V. Chen$^{3}$
\end{center}
\begin{center}
    \textit{\small $^1$Department of Physics and HK Institute of Quantum Science \& Technology,\\The University of Hong Kong, Pokfulam Road, Hong Kong, China\\
    $^2$Quantum Science Center of Guangdong-Hong Kong-Macau Great Bay Area, 3 Binlang Road, Shenzhen, China\\
    $^3$International Center for Quantum Materials, 
School of Physics, Peking University, Beijing 100871, China}
\end{center}

\setcounter{equation}{0}
\setcounter{figure}{0}
\setcounter{table}{0}
\setcounter{page}{1}
\makeatletter
\renewcommand{\theequation}{S\arabic{equation}}
\renewcommand{\thefigure}{S\arabic{figure}}
\renewcommand{\bibnumfmt}[1]{[S#1]}
\renewcommand{\citenumfont}[1]{S#1}

\section{Lattice vibration}
In this section, we derive the phonon Hamiltonian Eq.~(3) in the main-text.

We begin from the displacement length between site $i$ and $j$ as
\begin{align}
    |\Delta\mathbf{R}_{ij}|=|\mathbf{R}_{ij}|-|\mathbf{R}^0_{ij}|=\sqrt{(\mathbf{R}^0_{ij}+\mathbf{u}_{ij})^2}-|\mathbf{R}^0_{ij}|=|\mathbf{R}_{ij}^0|\sqrt{1+2\frac{\mathbf{R}_{ij}^0\cdot\mathbf{u}_{ij}}{|\mathbf{R}_{ij}^0|^2}+\frac{\mathbf{u}_{ij}^2}{|\mathbf{R}_{ij}^0|^2}}-|\mathbf{R}^0_{ij}|\approx\hat{\mathbf{R}}_{ij}^0\cdot\mathbf{u}_{ij}.
\end{align}
Therefore, to the lowest order, only in-plane displacement along bond $ij$ presents in the vibrational Hamiltonian, and by Hooke's law, we obtain Eq.~(3).

Next, we perform a Fourier transform for $\mathbf{u}_i$ as $\mathbf{u}_i=\frac{1}{\sqrt{N}}\sum_\mathbf{k}\mathbf{u}_{\mathbf{k}}e^{i\mathbf{k}\cdot\mathbf{R}_i}$ where $N$ is the number of the unit cell, and take it into Eq.~(3) for triangular lattice:
\begin{align}
    H_p(\mathbf{k})=\frac{1}{2M}|\mathbf{p}_{\mathbf{k}}|^2+\frac{1}{2}\mathbf{u}_{\mathbf{k}}^\dagger\mathbf{D}_p(\mathbf{k})\mathbf{u}_{\mathbf{k}}
\end{align}
where $\mathbf{D}_p(\mathbf{k})$ is the dynamical matrix for phonons as
\begin{align}
    \mathbf{D}_p(\mathbf{k})=M\omega_0^2\sum_i\begin{pmatrix}
    1-2\cos^2\phi_i\cos\mathbf{k}_i & -\sin2\phi_i\cos\mathbf{k}_i\\
    -\sin2\phi_i\cos\mathbf{k}_i & 1-2\sin^2\phi_i\cos\mathbf{k}_i
    \end{pmatrix}
\end{align}
where $\mathbf{k}_i=\mathbf{k}\cdot\boldsymbol{\delta}_i$ and $\phi_i=\{0,2\pi/3,4\pi/3\}$ is the angel for bond $\boldsymbol{\delta}_i=\{(1,0), (-1/2,\sqrt{3}/2), (-1/2,-\sqrt{3}/2)\}$ (with lattice constant $a=1$) for $i=1,2,3$ respectively. Then,
\begin{align}
    \mathbf{D}_p(\mathbf{k})=M\omega_0^2
\begin{pmatrix}
 3-\cos\frac{k_x}{2}\cos\frac{\sqrt{3}\eta_y}{2}-2 \cos k_x & \sqrt{3} \sin \frac{k_x}{2} \sin \frac{\sqrt{3} \eta_y}{2} \\
 \sqrt{3} \sin \frac{k_x}{2} \sin \frac{\sqrt{3} \eta_y}{2} & 3-3 \cos \frac{k_x}{2} \cos \frac{\sqrt{3} \eta_y}{2} \\
\end{pmatrix}
\end{align}

\section{Symmetry-adapted Magnetoelastic couplings}
In general, the spin Hamiltonian with nearest-neighbour symmetric interactions is written as
\begin{align}
    H_s=\sum_{\langle ij\rangle}\sum_{\alpha\beta}J_{ij}^{\alpha\beta}S_i^\alpha S_j^\beta,
\end{align}
where $J_{ij}^{\alpha\beta}$ is the exchange coupling between $\alpha$-component of $\mathbf{S}_i$ at site $\mathbf{R}_i$ and $\beta$-component of $\mathbf{S}_j$ at site $\mathbf{R}_j$. In the simplest form, the symmetric exchange depends on the bond length 
$|\mathbf{R}_{ij}|$ as 
\begin{align}
    J_{ij}^{\alpha\beta}=J_{ij}^{\alpha\beta}(|\mathbf{R}_{ij}|)=J_{ij}^{\alpha\beta}(|\mathbf{R}^0_{ij}
+\mathbf{u}_{ij}|)\approx J_{ij}^{\alpha\beta}(|\mathbf{R}^0_{ij}|)
+\frac{d J_{ij}^{\alpha\beta}}{dR}\hat{\mathbf{R}}^0_{ij}\cdot\mathbf{u}_{ij}.
\end{align}
A magnetoelastic interaction then emerges naturally from spin interactions as 
\begin{align}
    H_{me}=\sum_{\langle ij\rangle}\sum_{\alpha\beta}K_{ij}^{\alpha\beta} 
    S_{i}^\alpha S_{j}^\beta(\hat{\mathbf{R}}_{ij}^0\cdot\mathbf{u}_{ij}),\quad \text{with}\quad K^{\alpha\beta}_{ij}=\frac{d J_{ij}^{\alpha\beta}(R)}{dR}.
\end{align}
For the triangular lattice with D$_3$ symmetry, the symmetry-adapted form [46] of $J$-tensor is
\begin{equation}
    J_{ij}=\begin{pmatrix}
        J+2J_{\pm}\cos 2\phi_{ij} & 2J_{\pm}\sin 2\phi_{ij} & 
        J_{z\pm}\sin 2\phi_{ij}\\2J_{\pm}\sin 2\phi_{ij} & J
        -2J_{\pm}\cos 2\phi_{ij} & J_{z\pm}\cos 2\phi_{ij}\\
        J_{z\pm}\sin 2\phi_{ij} & J_{z\pm}\cos 2\phi_{ij} & \Delta J
    \end{pmatrix}
\end{equation}
where $\phi_{ij}=\{0, 2\pi/3,4\pi/3\}$ is $\langle ij \rangle$-bond angle 
with respect to the $x$-axis. 

Therefore, the corresponding symmetry-adapted form of $K$-tensor is
\begin{equation}
    K_{ij}=\begin{pmatrix}
        K_0+K_1\cos 2\phi_{ij} & K_1\sin 2\phi_{ij} & K_2\sin 2\phi_{ij}\\K_1\sin 2\phi_{ij} & K_0-K_1\cos 2\phi_{ij} & K_2\cos 2\phi_{ij}\\K_2\sin 2\phi_{ij} & K_2\cos 2\phi_{ij} & K_3
    \end{pmatrix},
\end{equation}
with 4 independent magnetoelastic coupling constants. In the case of D$_{3h}$, the symmetry further requires $J_{z\pm}=0$ as well as $K_2=0$.

More generally, given that the magnetoelastic coupling is a rank-4 tensor, the magnetoelastic energy density can be written as
\begin{align}
    h_{me}=\kappa_{\alpha\beta\gamma\delta}m^\alpha m^\beta \epsilon^{\gamma\delta}, 
\end{align}
where $m^\alpha$ is the $\alpha$-component of the magnetization cosine, $\epsilon^{\gamma\delta}=\frac{1}{2}\left(\frac{\partial u^\gamma}{\partial r_\delta}+\frac{\partial u^\delta}{\partial r_\gamma}\right)$ is the strain tensor for lattice displacement $\mathbf{u}(\mathbf{r})$, and $\kappa_{\alpha\beta\gamma\delta}$ is the corresponding magnetoelastic coupling constant.

Obvisouly, $\kappa_{\alpha\beta\gamma\delta}=\kappa_{\beta\alpha\gamma\delta}=\kappa_{\alpha\beta\delta\gamma}$. Besides, a lattice symmetry $g$ with the matrix representation $R_g$ requires $\kappa_{\alpha'\beta'\gamma'\delta'}=R_g^{\alpha\alpha'}R_g^{\beta\beta'}R_g^{\gamma\gamma'}R_g^{\delta\delta'}\kappa_{\alpha\beta\gamma\delta}$. These constraints determine the symmetry-adapted form of the magnetoelastic tensor.

In the two-dimensional lattice model, we neglect $u^z$ and the derivative of $z$, and consider only in-plane displacements along bonds due to Eq.~(S1). In other words, the only non-zero strain is
\begin{align}
    \epsilon^{x'x'}\approx \frac{\hat{\mathbf{R}}_{ij}^0\cdot\mathbf{u}_{ij}}{|\mathbf{R}_{ij}^0|},\quad \text{with} \quad \hat{\mathbf{x}}'=\hat{\mathbf{R}}_{ij}^0.
\end{align}
In the global coordinate, $\epsilon^{xx}=\epsilon^{x'x'}\cos^2\phi_{ij},\ \epsilon^{yy}=\epsilon^{x'x'}\sin^2\phi_{ij},\ \epsilon^{xy}=\epsilon^{yx}=\epsilon^{x'x'}\cos\phi_{ij}\sin\phi_{ij}$ for bond $ij$. Then the magnetoelastic Hamiltonian can be written as
\begin{align}
    H_{me}&=vh_{me}\nonumber\\
    &=\sum_{\langle ij\rangle}\sum_{\alpha\beta}\frac{v}{|\mathbf{R}_{ij}^0|S^2}(\kappa_{\alpha\beta xx}\cos^2\phi_{ij}+\kappa_{\alpha\beta yy}\sin^2\phi_{ij}+\kappa_{\alpha\beta xy}\sin\phi_{ij}\cos\phi_{ij}+\kappa_{\alpha\beta yx}\cos\phi_{ij}\sin\phi_{ij})S_i^\alpha S_j^\beta(\hat{\mathbf{R}}_{ij}^0\cdot\mathbf{u}_{ij})\nonumber\\
    &=\sum_{\langle ij\rangle}\sum_{\alpha\beta}\frac{v}{|\mathbf{R}_{ij}^0|S^2}\left(\frac{\kappa_{\alpha\beta xx}+\kappa_{\alpha\beta yy}}{2}+\frac{\kappa_{\alpha\beta xx}-\kappa_{\alpha\beta yy}}{2}\cos2\phi_{ij}+\frac{\kappa_{\alpha\beta xy}+\kappa_{\alpha\beta yx}}{2}\sin2\phi_{ij}\right)S_i^\alpha S_j^\beta(\hat{\mathbf{R}}_{ij}^0\cdot\mathbf{u}_{ij}),
\end{align}
which is consistent with Eq.~(S7) with $K^{\alpha\beta}_{ij}\equiv \frac{v}{|\mathbf{R}_{ij}^0|S^2}\left(\frac{\kappa_{\alpha\beta xx}+\kappa_{\alpha\beta yy}}{2}+\frac{\kappa_{\alpha\beta xx}-\kappa_{\alpha\beta yy}}{2}\cos2\phi_{ij}+\frac{\kappa_{\alpha\beta xy}+\kappa_{\alpha\beta yx}}{2}\sin2\phi_{ij}\right)$, where $v$ is the unit cell volume. 

In the main-text, we determine the independent coupling constants $\kappa_{\alpha\beta\gamma\delta}$ for D$_3$ and other lattice symmetries of hexagonal crystal family with the help of TENSOR [51] in Bilbao Crystallographic Server. For example, there are only 4 independent coefficients $K_{11}, K_{12}, K_{13}, K_{14}$ for D$_3$ with $\kappa_{xxxx}=\kappa_{yyyy}=K_{11}$, $\kappa_{xxyy}=\kappa_{yyxx}=K_{12}$, $\kappa_{zzxx}=\kappa_{zzyy}=K_{13}$, $\kappa_{xzxy}=\kappa_{zxxy}=\kappa_{xzyx}=\kappa_{zxyx}=K_{14}$, $\kappa_{yzxx}=-\kappa_{yzyy}=K_{14}$, and $\kappa_{xyxy}=\kappa_{xyyx}=\kappa_{yxxy}=\kappa_{yxyx}=\frac{1}{2}(K_{11}-K_{12})$. One can check that, with $K_0\equiv\frac{v}{2|\mathbf{R}_{ij}^0|S^2}(K_{11}+K_{12})$, $K_1\equiv\frac{v}{2|\mathbf{R}_{ij}^0|S^2}(K_{11}-K_{12})$, $K_2\equiv\frac{v}{2|\mathbf{R}_{ij}^0|S^2}K_{14}$, and $K_3\equiv \frac{v}{|\mathbf{R}_{ij}^0|S^2}K_{13}$, Eq.~(S12) is consistent with Eq.~(S9). For D$_{3h}$, TENSOR gives $K_{14}=0$ so that $K_2=0$.

\section{Magnetoelastic Hamiltonian}
In this section, we derive the expression for $H_{c}(\mathbf{k})$ of Eq.~(9) from $H_{me}$. To couple spin-waves with phonons in the lowest order, we only need to consider $K_{ij}^{x(z),y}$ and $K_{ij}^{y,x(z)}$ terms in $H_{me}$:
\begin{align}
    H_{me}=SK_1\sum_{\langle ij\rangle}\sin\left(2\phi_{ij}\right)(S_{i}^x+S_{j}^x)(\hat{\mathbf{R}}_{ij}^0\cdot\mathbf{u}_{ij})+SK_2\sum_{\langle ij\rangle}\cos\left(2\phi_{ij}\right)(S_{i}^z+S_{j}^z)(\hat{\mathbf{R}}_{ij}^0\cdot\mathbf{u}_{ij}).
\end{align}
In the $\mathbf{k}$-space Holstein-Primakoff representation,
\begin{align}
    H_{me}(\mathbf{k})&=(a_\mathbf{k}^\dagger,a_{-\mathbf{k}})\sum_i\left[\sqrt{\frac{S^3}{2}}\begin{pmatrix}
        K_1\sin 2\phi_i+iK_2(1+\cos 2\phi_i) & K_1(1-\cos 2\phi_i)+iK_2 \sin 2\phi_i\\
        -K_1\sin 2\phi_i+iK_2(1+\cos 2\phi_i) & -K_1(1-\cos 2\phi_i)+iK_2 \sin 2\phi_i
    \end{pmatrix}\sin\mathbf{k}_i\right]\begin{pmatrix}
        u^x_\mathbf{k}\\u^y_\mathbf{k}
    \end{pmatrix}+h.c.\nonumber\\
    &=(a_\mathbf{k}^\dagger,a_{-\mathbf{k}})H_c(\mathbf{k})\begin{pmatrix}
        u^x_\mathbf{k}\\u^y_\mathbf{k}
    \end{pmatrix}+h.c.,
\end{align}
as written in Eq.~(9).

\section{Emergent gauge fields}
Under the ``phase coordinate'' $\mathbf{Y}_\mathbf{k}=(\mathbf{Q}_{\mathbf{k}}; 
\mathbf{P}^\dagger_{\mathbf{k}})^T$, the equations of motion $i\dot{\mathbf{Y}}_\mathbf{k}=[\mathbf{Y}_\mathbf{k},H]$ gives rise to
\begin{align}
    \begin{pmatrix}
        \dot{\mathbf{Q}}_\mathbf{k}\\ \dot{\mathbf{P}}^\dagger_{\mathbf{k}}
    \end{pmatrix}=\begin{pmatrix}
        \mathbf{A}_\mathbf{k}^\dagger & \frac{1}{M}I_{3}\\
        -\mathbf{D}_\mathbf{k} & -\mathbf{A}_\mathbf{k}
    \end{pmatrix}\begin{pmatrix}
        \mathbf{Q}_\mathbf{k}\\ \mathbf{P}^\dagger_{\mathbf{k}},
    \end{pmatrix}\Rightarrow\left\{\begin{array}{cc}
       \mathbf{P}^\dagger_{\mathbf{k}}&=M(\dot{\mathbf{Q}_\mathbf{k}}-\mathbf{A}_\mathbf{k}^\dagger\mathbf{Q}_\mathbf{k})  \\
        \mathbf{D}_\mathbf{k}\mathbf{Q}_\mathbf{k}&=-\dot{\mathbf{P}}^\dagger_{\mathbf{k}}-\mathbf{A}_\mathbf{k}\mathbf{P}^\dagger_{\mathbf{k}} 
    \end{array}\right.\label{EoM}
\end{align}

Further, we can obtain the dynamical equation for $\mathbf{Q}_\mathbf{k}$ from Eq.~\eqref{EoM} as
\begin{align}
\ddot{\mathbf{Q}}_\mathbf{k}=-\Tilde{\mathbf{D}}_{\mathbf{k}}\mathbf{Q}_\mathbf{k}+\Gamma_\mathbf{k}\dot{\mathbf{Q}}_\mathbf{k}\label{EoMQ}
\end{align}
with $\tilde{\mathbf{D}}_{\mathbf{k}}=\frac{1}{M}\mathbf{D}_\mathbf{k}-\mathbf{A}_\mathbf{k}\mathbf{A}_\mathbf{k}^\dagger$ and $\Gamma_\mathbf{k}=\mathbf{A}^\dagger_\mathbf{k}-\mathbf{A}_\mathbf{k}\equiv -2\tilde{\mathbf{A}}_{\mathbf{k}}$.

We can then define an artificial $\mathbf{k}$-space magnetic field $\mathbf{B}_\mathbf{k}=(B_\mathbf{k}^x, B_\mathbf{k}^y, 0)$ from $\tilde{\mathbf{A}}_{\mathbf{k}}$, where
\begin{align}
\tilde{\mathbf{A}}_{\mathbf{k}}=\frac{1}{2M}\begin{pmatrix}
    0 & 0 & B_\mathbf{k}^y\\
    0 & 0 & -B_\mathbf{k}^x\\
    -B_\mathbf{k}^y & B_\mathbf{k}^x & 0
\end{pmatrix},\quad \text{with}\quad  \left\{\begin{array}{ll}
    B_\mathbf{k}^x=&MK_2\frac{\sqrt{2S^3}}{2r_\mathbf{k}}\sum_i\sin 2\phi_i\sin\mathbf{k}_i,\\
         B_\mathbf{k}^y=& -MK_2\frac{\sqrt{2S^3}}{2r_\mathbf{k}}\sum_i(1+\cos 2\phi_i)\sin\mathbf{k}_i,
    \end{array}\right.
\end{align}
and one can check that the equation of motion Eq.~\eqref{EoMQ} is the same as that of $\tilde{H}=\frac{1}{2M}\sum_{\mathbf{k}}|\mathbf{P}_{-\mathbf{k}}-\frac{1}{2}\mathbf{B}_\mathbf{k}\times\mathbf{Q}_\mathbf{k}|^2+\frac{M}{2}\mathbf{Q}_\mathbf{k}^\dagger\tilde{\mathbf{D}}_{\mathbf{k}}\mathbf{Q}_\mathbf{k}$, which describes a phononic system with spin Raman interaction although the magnetic field now is a dynamical one with a dependence on $\mathbf{k}$.

\section{Berry physics and topological classification}
The eigen-equation for this magnetoelastic coupled system under the basis $\mathbf{X}_\mathbf{k}$ is written as
\begin{align}
    \mathcal{J} H_{\mathbf{k}}|\psi_{n\mathbf{k}}\rangle=E_{n\mathbf{k}}|\psi_{n\mathbf{k}}\rangle,
\end{align}
where $\ket{\psi_{n\mathbf{k}}}$ is the Bloch wavefunction for $n$-th band, $H_\mathbf{k}$ is the Hamiltonian under the basis $\mathbf{X}_\mathbf{k}$, and thus the proper Lagrangian should be
\begin{align}
    \mathcal{L}=i\frac{d}{dt}-\mathcal{J} H_\mathbf{k}.
\end{align}
Following the derivation for the electronic system, one can obtain the proper Berry connection $\boldsymbol{\mathcal{A}}_{n\mathbf{k}}$ and Berry curvature $\mathbf{\Omega}_{n\mathbf{k}}$ for magneto-phonons as
\begin{align}
    \boldsymbol{\mathcal{A}}_{n\mathbf{k}}=i\left<\psi_{n\mathbf{k}}\right|\mathcal{J}\partial_{\mathbf{k}}\left|\psi_{n\mathbf{k}}\right>,\quad \mathbf{\Omega}_{n\mathbf{k}}=i\left<\nabla_\mathbf{k}\psi_{n\mathbf{k}}\right|\mathcal{J}\times\left|\nabla_\mathbf{k}\psi_{n\mathbf{k}}\right>.\label{BC}
\end{align}
In the main text, we show that magnetoelastic couplings in the triangular lattice with D$_3$ symmetry lead to non-zero integer band Chern numbers.

To understand the effects of the lattice symmetry, we study the topological classification of the equivalent phononic system
\begin{align}
    H=\frac{1}{2}\sum_{\mathbf{k}}(\mathbf{Q}^\dagger_{\mathbf{k}},\mathbf{P}_{\mathbf{k}}) \begin{pmatrix}
        \mathbf{D}_\mathbf{k} & \mathbf{A}_\mathbf{k}\\
        \mathbf{A}_\mathbf{k}^\dagger & \frac{I_3}{M}
    \end{pmatrix}\begin{pmatrix}
        \mathbf{Q}_{\mathbf{k}}\\\mathbf{P}^\dagger_{\mathbf{k}}
    \end{pmatrix}\quad \text{with}\quad \mathbf{D}_\mathbf{k}=\begin{pmatrix}
        \mathbf{D}_p(\mathbf{k}) & \mathbf{D}_{me}(\mathbf{k})\\
        \mathbf{D}^T_{me}(\mathbf{k}) & M\varepsilon^2_{m}(\mathbf{k})
    \end{pmatrix}.
\end{align}

Since $i\begin{pmatrix}
        \dot{\mathbf{Q}}_\mathbf{k}\\ \dot{\mathbf{P}}^\dagger_{\mathbf{k}}
    \end{pmatrix}=i\begin{pmatrix}
        \mathbf{A}_\mathbf{k}^\dagger & \frac{1}{M}I_{3}\\
        -\mathbf{D}_\mathbf{k} & -\mathbf{A}_\mathbf{k}
    \end{pmatrix}\begin{pmatrix}
        \mathbf{Q}_\mathbf{k}\\ \mathbf{P}^\dagger_{\mathbf{k}},
    \end{pmatrix}$, with the spirit of classification on topological phonons, we define $\bar{H}_\mathbf{k}\equiv i\begin{pmatrix}
        \mathbf{A}_\mathbf{k}^\dagger & \frac{1}{M}I_{3}\\
        -\mathbf{D}_\mathbf{k} & -\mathbf{A}_\mathbf{k}
    \end{pmatrix}$ and consider the ``time-reversal'' symmetry $\mathcal{T}$, particle-hole symmetry $\mathcal{C}$, and chiral symmetry $\mathcal{S}$ of $\bar{H}_\mathbf{k}$. Here, we use the same notation and definition of these symmetries as those in Ref.~[45]:\\
(1) The system is $\mathcal{T}$ symmetric if an antiunitary matrix $U_\mathcal{T}$ transform $\bar{H}_\mathbf{k}$ as
\begin{align}
    U_\mathcal{T}\bar{H}_\mathbf{k}-\bar{H}_{-\mathbf{k}}U_\mathcal{T}=0,\quad\text{with }U_\mathcal{T}^2=\pm 1;
\end{align}
(2) The system is $\mathcal{C}$ symmetric, if an antiunitary matrix $U_\mathcal{C}$ satisfy 
\begin{align}
    U_\mathcal{C}\bar{H}_\mathbf{k}+\bar{H}_{-\mathbf{k}}U_\mathcal{C}=0,\quad\text{with }U_\mathcal{C}^2=\pm 1;
\end{align}
(3) If the system is $\mathcal{S}$ symmetric, we demand an unitary matrix $U_\mathcal{S}$ so that
\begin{align}
    U_\mathcal{S}\bar{H}_\mathbf{k}+\bar{H}_{\mathbf{k}}U_\mathcal{S}=0,\quad\text{with }U_\mathcal{S}^2=1.
\end{align}

Since $\mathbf{A}_\mathbf{k}=\mathbf{A}^*_\mathbf{k}$, $\mathbf{D}_\mathbf{k}=\mathbf{D}^*_{\mathbf{k}}$, and $W\mathbf{A}_\mathbf{k}=\mathbf{A}_{-\mathbf{k}}W$, $W\mathbf{D}_\mathbf{k}=\mathbf{D}_{-\mathbf{k}}W$, with $W=\begin{pmatrix}
    1 & 0 & 0\\0 & 1 & 0\\0 & 0 & -1
\end{pmatrix}$. It is easy to find the particle-hole symmetry can always be achieved by $U_\mathcal{C}=\begin{pmatrix}
    W & 0\\0&W
\end{pmatrix}\mathcal{K}$ with $U_\mathcal{C}^2=+1$, where $\mathcal{K}$ is the complex conjugation operator, so that
\begin{align}
    U_\mathcal{C}\bar{H}_\mathbf{k}=-i\begin{pmatrix}
    W & 0\\0&W
\end{pmatrix}\mathcal{K}\begin{pmatrix}
        \mathbf{A}_\mathbf{k}^\dagger & \frac{1}{M}I_{3}\\
        -\mathbf{D}_\mathbf{k} & -\mathbf{A}_\mathbf{k}
    \end{pmatrix}=-i\begin{pmatrix}
        W\mathbf{A}_\mathbf{k}^\dagger & \frac{1}{M}W\\
        -W\mathbf{D}_\mathbf{k} & -W\mathbf{A}_\mathbf{k}
    \end{pmatrix}\mathcal{K}=-i\begin{pmatrix}
        \mathbf{A}_{-\mathbf{k}}^\dagger W & \frac{1}{M}W\\
        -\mathbf{D}_{-\mathbf{k}}W & -\mathbf{A}_{-\mathbf{k}}W
    \end{pmatrix}\mathcal{K}=-\bar{H}_{-\mathbf{k}}U_\mathcal{C}
\end{align}

If $K_2=0$ so that $\mathbf{A}_\mathbf{k}=0$, then $U_\mathcal{T}=\begin{pmatrix}
    W & 0\\0 & -W
\end{pmatrix}\mathcal{K}$ with $U_\mathcal{T}^2=+1$, and 
\begin{align}
    U_\mathcal{T}\bar{H}_\mathbf{k}=-i\begin{pmatrix}
    W & 0\\0 & -W
\end{pmatrix}\mathcal{K}\begin{pmatrix}
        0 & \frac{1}{M}I_{3}\\
        -\mathbf{D}_\mathbf{k} & 0
    \end{pmatrix}=-i\begin{pmatrix}
        0 & \frac{1}{M}W\\
        W\mathbf{D}_\mathbf{k} & 0
    \end{pmatrix}\mathcal{K}=i\begin{pmatrix}
        0 & -\frac{1}{M}W\\
        -\mathbf{D}_{-\mathbf{k}}W & 0
    \end{pmatrix}\mathcal{K}=\bar{H}_{-\mathbf{k}}U_\mathcal{T}
\end{align}
Consequently, $U_\mathcal{S}=U_\mathcal{T}U_\mathcal{C}$. Therefore, the system belongs to the BDI class with trivial topology in two dimensions. Whereas if $K_1\neq0$ and $K_2\neq0$, though $D_{me}(\mathbf{k})$ contains magnetoelastic effects, $\mathcal{C}$-symmetry is the only symmetry, and thus the system belongs to D class characterized by an integer number $(\mathbb{Z})$ in two dimensions.

Below in Fig.~\ref{fig:BDI} we plot the dispersion with the color showing the chirality for $K_2=0$, and we find the bands are modified by the magnetoelastic effects but are not fully gapped and the chirality is zero for any $\mathbf{k}$.
\begin{figure}[ht!]
\subfigure[Band dispersion with $K_1=6$ meV/\AA and $K_2=0$. The black solid (dashed) curves are the magnon (phonon) dispersion with $K_1=K_2=0$.]{\includegraphics[width=0.56\textwidth]{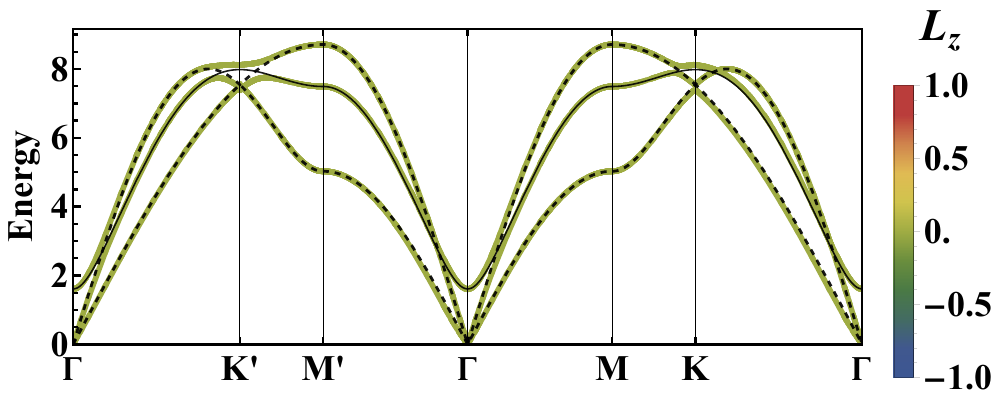}}\subfigure[Phonon angular momentum with $K_1=6$ meV/\AA and $K_2=0$.]{\includegraphics[width=0.415\textwidth]{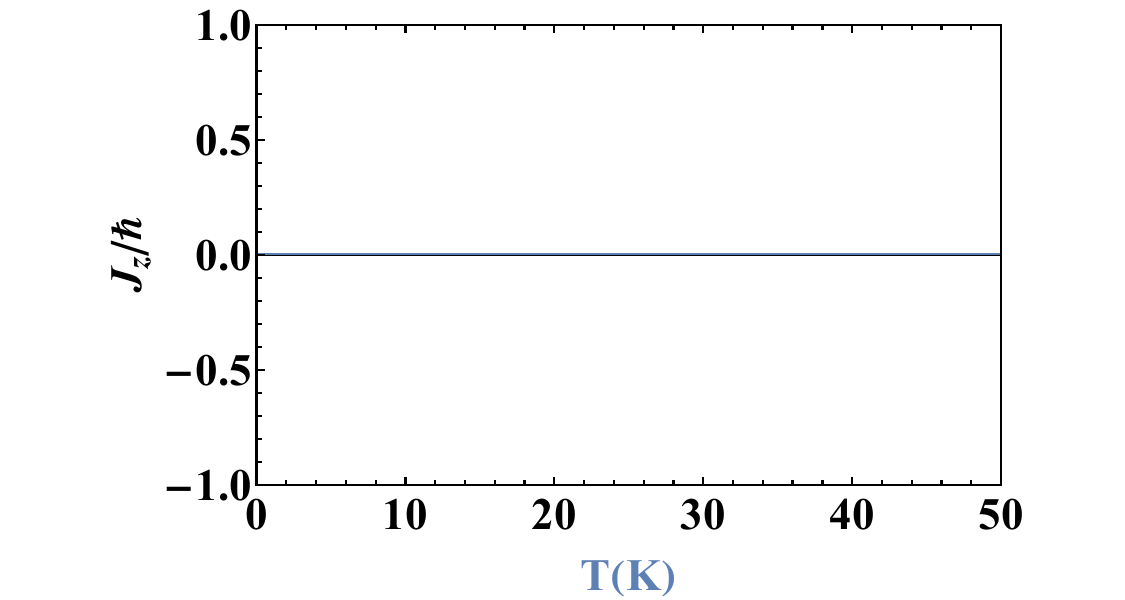}}
\caption{(a) The band dispersion and (b) phonon angular momentum with $K_1=6$ meV/\AA\ but $K_2=0$. Other parameters are the same with those in Fig.~1(c). The black solid (dashed) line is the magnon (phonon) dispersion without MEC.} \label{fig:BDI}
\end{figure}

In principle, this approach to classifying the topology and chirality of the hybrid excitations can be generally used in any magnetoelastic coupled systems or magnon-phonon interacted systems from other mechanisms.
\\
\\
\\
\\
\\
\section{More general bond-dependent spin Hamiltonian}
As shown in Eq.~(S8), the bond-dependent interaction $J_{z\pm}$ (and $J_{\pm}$) is generally allowed under $D_{3h} (D_3)$ symmetry. However, since the key element to the chiral and topological excitation in this magnon-phonon coupled system is the magnetoelastic couplings, introducing these non-zero bond-dependent interactions will only modify the magnon dispersion but not change the main conclusions in the main text. Below in Fig.~\ref{fig:S2}, we show a general bond-dependent spin Hamiltonian with ${S=3/2}$, ${J=-1.3}$ meV, ${\Delta=0.5}$, ${J_{\pm}=-0.4}$ meV, 
${J_{z\pm}=0.2}$ meV, ${B_\text{eff}=1}$ meV, ${\hbar\omega_0=7.7}$ meV, 
${M = 56}$ u, ${\kappa_1=6}$ meV/\AA, and (a) $\kappa_2=6$ meV/\AA\ (b) $\kappa_2=0$ meV/\AA. Indeed, non-zero phonon angular momentum only occurs when $\kappa_2\neq 0$.
\begin{figure}[ht!]
\subfigure[Band dispersion with $K_1=K_2=6$ meV/\AA.]{\includegraphics[width=0.48\textwidth]{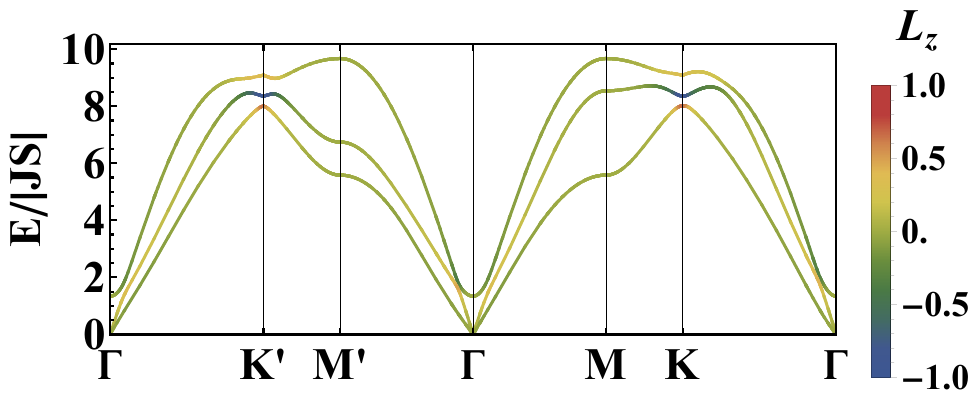}}\subfigure[Band dispersion with $K_1=6$ meV/\AA\ but $K_2=0$.]{\includegraphics[width=0.48\textwidth]{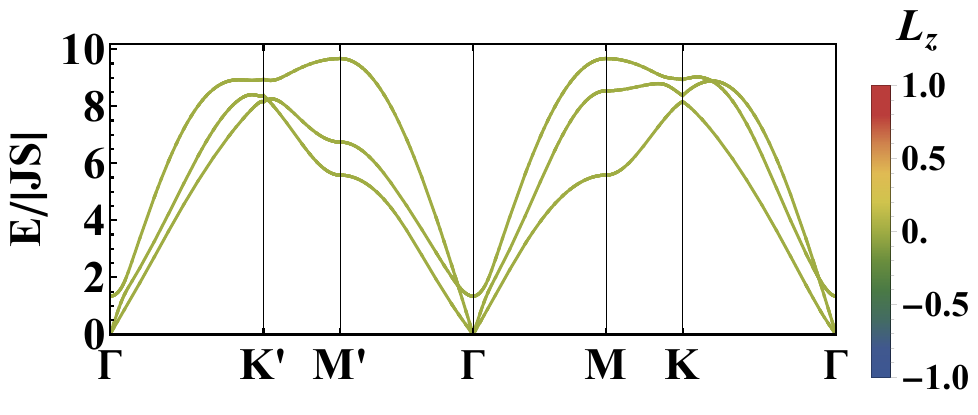}}
\caption{The band dispersion of a general bond-dependent spin Hamiltonian with $J_{\pm}\neq 0$ and $J_{z\pm}\neq 0$. The $\mathbf{k}$-resolve phonon angular momentum is indicated by color.} \label{fig:S2}
\end{figure}
\end{document}